%% file: main_arxiv.tex
\documentclass[a4paper,
11pt
]{article}

\usepackage[english]{babel}
\usepackage[a4paper,left=2.5cm,right=2.5cm,top=2.5cm,bottom=2.5cm]{geometry}
\usepackage[colorlinks=true, urlcolor=blue]{hyperref}
\usepackage{bm}
\usepackage{authblk}
\usepackage{siunitx}
\usepackage[ruled,vlined]{algorithm2e}
\usepackage{algorithmic}
\usepackage{caption}
\usepackage{subcaption}
\usepackage{amssymb} 
\usepackage[dvipsnames]{xcolor}
\usepackage{amsmath, amssymb} 
\usepackage{booktabs}
\usepackage{rotating}
\usepackage{tikz}
\usepackage{colortbl} 
\usepackage{multirow}
\usepackage{array}
\usepackage{url}
\usepackage[square,numbers]{natbib}
\usepackage{enumitem} 
\usepackage{lineno}

\graphicspath{ {./images/} }

\title{{ 
Data-driven reduced order model for residence time distribution analysis of an industrial-scale continuous casting tundish
}}
\author[1,4]{Harshith~Gowrachari\footnote{hgowrach@sissa.it, harshith.gowrachari@onera.fr}}
\author[3]{Mattia~Giuseppe~Barra\footnote{m.barra@danieli.com}}
\author[2]{Giovanni~Stabile\footnote{giovanni.stabile@santannapisa.it}}
\author[3]{Gianluca~Bazzaro\footnote{g.bazzaro@danieli.it}}
\author[1]{Gianluigi~Rozza\footnote{grozza@sissa.it}}

\affil[1]{\small Mathematics Area, mathLab, International School for Advanced Studies\\

Via Bonomea 265, 34136, Trieste, Italy.}

\affil[2]{\small Biorobotics Institute, Sant’Anna School of Advanced Studies\\

V.le R. Piaggio 34, 56025, Pontedera, Pisa, Italy.}

\affil[3]{\small Danieli Research Center, Danieli $\&$ C. S.p.A., Via Nazionale 41\\

 33042, Buttrio, Province of Udine, Italy.}

\affil[4]{Present address: DTIS, ONERA -- The French Aerospace Lab\\

2 Avenue Marc Pélegrin, 31400, Toulouse, France.}

\date{} % Leave empty to omit a date

\begin{document}
\maketitle
    \input{sections/abstract}
    \input{sections/intro}

    \input{sections/FOM}
    \input{sections/ROM}

    \input{sections/domain}
    \input{sections/numerical_experiments}
    \input{sections/conclusion}
    \input{sections/ackno}
\bibliographystyle{abbrvnat}
\bibliography{main_arxiv.bbl}

\end{document}

%% file: sections/abstract.tex
\begin{abstract}
The continuous casting tundish plays a critical role as a metallurgical reactor in the continuous casting process, with its flow characteristics serving as a key parameter in the production of high-quality steel. These characteristics are typically assessed through residence time distribution (RTD) curves. This study examines the flow behaviour in a single-strand continuous casting tundish through a combination of numerical simulations and experimental validation. Steady-state full order model (FOM) simulations are performed under both isothermal and non-isothermal conditions to evaluate the influence of thermal buoyancy on the velocity field, which is found to be negligible. The resulting flow fields are used to initialize transient tracer transport simulations for determining the RTD and flow volume partitioning. Subsequently, a data-driven reduced order model (ROM) is developed to predict the RTD response. Comparison of RTD curves obtained from experiments, FOM, and ROM shows excellent agreement, with the ROM accurately capturing the key flow characteristics at a fraction of the computational cost. These results highlight the potential of ROM techniques for efficient real-time analysis, design, and optimization of tundish operations in metallurgical processes. \\


\end{abstract}

%% file: sections/intro.tex
\section{Introduction}\label{sec:intro}

\begin{figure}[ht!]
    \centering
    \includegraphics[width=0.65\linewidth]{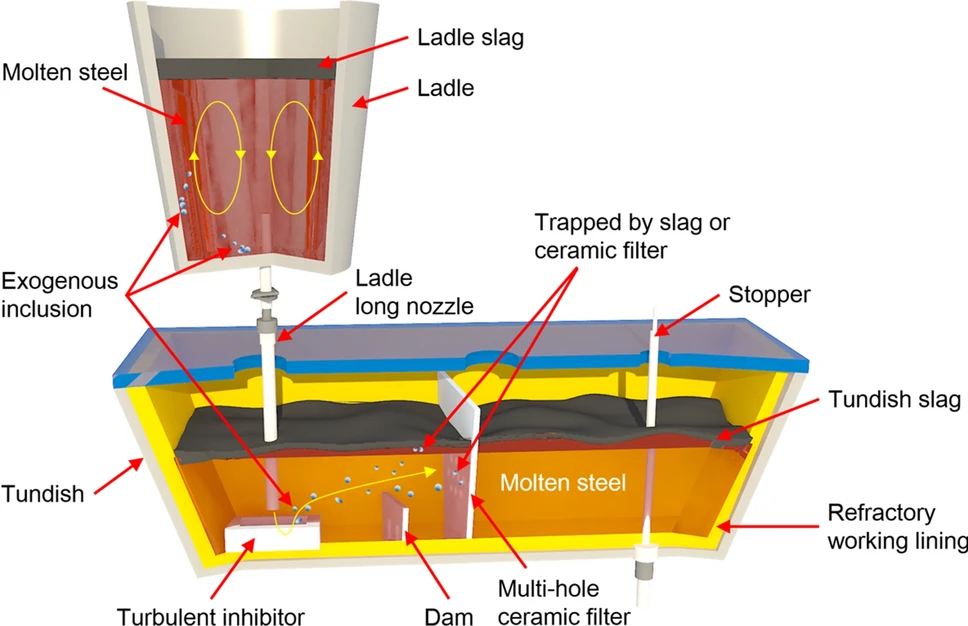}
    \caption{Schematic of a single-strand tundish (taken from \cite{Wang2021}), illustrating key components and flow features. The diagram depicts multiphase interactions involving molten steel, slag, and inclusion flotation. Flow control devices, such as the impact pad (turbulence inhibitor) and dam, are also shown, highlighting their role in enhancing steel quality and casting performance.}
    \label{fig:Tundish}
\end{figure}

{ Continuous casting is the dominant method for producing semi-finished steel products such as billets, blooms, and slabs, accounting for approximately 96\% of global steel production \cite{worldsteel2024}. A key component of this process is the tundish (see Figure~\ref{fig:Tundish}), a refractory-lined vessel that transfers molten steel from the ladle to the mold. While traditionally considered a passive buffer between secondary metallurgical operations and casting, the tundish is now recognized as an active metallurgical reactor. It plays a crucial role in controlling flow dynamics and promoting the removal of non-metallic inclusions, which strongly influence the cleanliness and final quality of steel \cite{Zhang2003, Tkadlekov2020, Ren2022}. Achieving high product quality, therefore, requires accurate prediction and control of flow behavior within the tundish, motivating the development of efficient predictive models \cite{Zhang2020}.}

A commonly employed method for characterizing molten steel flow in tundishes is the stimulus–response technique \cite{Wang2022}. This involves injecting a tracer at the tundish inlet and monitoring its concentration at the outlet over time, yielding the residence time distribution (RTD) curve \cite{Sahai2007}. Step and pulse tracer inputs are typically used. In multi-strand tundishes, the global RTD is computed as a flow-weighted average of outlet-specific RTDs, accounting for unequal flow splits. \\
{

RTD analysis offers insights into flow behavior, including the detection of preferential flow paths, shortcuts that allow molten steel to exit with minimal residence time, and limiting the removal of non-metallic inclusions. These appear as early peaks in the RTD curve and are detrimental to steel cleanliness. For RTD-based volume partitioning, the widely adopted combined model for a single-strand tundish proposed by Sahai and Emi \cite{Sahai1996} is employed. According to this model, the tundish flow domain is divided into three distinct regions: plug flow, fully mixed flow, and dead flow. The plug flow and fully-mixed flow regions together constitute the active zone. In this framework, the tundish is represented as an assembly of idealized reactor elements. Subsequently, Sahai and Ahuja \cite{Sahai1986} refined the original formulation and proposed an improved combined model, expressed as follows:

\begin{equation}
\frac{V_{\text {dead }}}{V}=1-\frac{T_{\text {avg }}}{T_\text{th}},
\label{eqn:deadVolume}
\end{equation}

\begin{equation}
\frac{V_{\text {plug }}}{V}=\frac{\frac{1}{2} (T_{\min } + T_{\max})}{T_\text{th}} = \frac{T_\text{plug}}{T_\text{th}},
\label{eqn:plugVolume}
\end{equation}

\begin{equation}
\frac{V_\text{mix}}{V} = 1 - \frac{V_{\text {dead }}}{V} - \frac{V_{\text {plug }}}{V},
\label{eqn:mixVolume}
\end{equation}

\begin{equation}
    T_\text{th} = \frac{V}{Q}.
    \label{eqn:Tth}
\end{equation}

\noindent where \(T_{\text{avg}}\) denotes the average residence time, defined as the time required for 50\% of the injected tracer to leave the tundish. The theoretical residence time, \(T_{\text{th}}\) (\ref{eqn:Tth}), is given by the ratio of the tundish volume \(V\) to the volumetric flow rate \(Q\). The parameter \(T_{\text{min}}\) represents the time at which the tracer concentration first attains \(1 \times 10^{-4}\), while \(T_{\max}\) corresponds to the time at which the peak tracer concentration is observed. \\

The dead volume fraction (\ref{eqn:deadVolume}) characterizes the portion of the tundish where molten steel resides for a prolonged period, typically greater than twice the theoretical residence time. The plug volume fraction (\ref{eqn:plugVolume}) is defined as the ratio of the mean time required to reach the maximum outlet concentration to the theoretical residence time, representing the fraction of the active volume exhibiting near plug-flow behavior. The mixing volume fraction (\ref{eqn:mixVolume}) corresponds to the portion of the active tundish volume in which the molten steel undergoes effective homogenization.
} \\

RTD curves can be derived from both numerical and physical simulations. Physical modelling is often performed using full or reduced-scale water models, with water representing molten steel and a NaCl solution serving as the tracer \cite{Sahai2007, Dinda2024}. {{{The accuracy of RTD measurements is strongly influenced by the tracer injection methodology and tracer properties, including concentration, density, and injection volume \cite{Liu2022, Damle1995, Brard2020}. Although lower tracer dosages help minimize flow disturbances, the injected concentration must remain within the detection limits of the measurement system \cite{Chen2012}. \\

In addition to tracer-related effects, tundish flow modifiers such as impact pads, dams, and ceramic filters significantly alter the internal flow structure and circulation characteristics, thereby influencing residence time distribution and dead volume within the tundish \cite{anapagaddi2013exploring, Anapagaddi2014, Bensouici2009, Tkadlekov2020}. Consequently, accurate numerical prediction of RTD behavior depends strongly on the turbulence model used to represent the flow dynamics. Numerical validation is typically performed by comparing simulation results with experimental data from water models \cite{Liu2022}}} \\

Reduced order models (ROMs) offer computationally efficient alternatives to high-fidelity numerical simulations, mitigating their high computational cost \cite{rozza2022advanced}. For instance, evaluating RTD curves across multiple operating parameters using full-order models (FOMs) is computationally expensive, motivating for the ROMs capable of capturing the dominant dynamics across a range of parameters. In ROM construction, once a low-dimensional approximation subspace of the FOM solution manifold is obtained, the system dynamics must be evolved within this subspace. This is typically achieved by solving for modal coefficients or latent variables, using either intrusive or non-intrusive approaches. \\

In the intrusive approach, the discretized governing equations are projected onto the reduced subspace, yielding a system of reduced order ordinary differential equations (ODEs). This projection based approach is commonly referred to as Galerkin or Petrov–Galerkin ROM, is inherently physics-based \cite{Benner2015, stabile2018finite, stabile2019reduced, lee2020model, romor2023non}. However, this approach requires full access to the underlying equations and the internal data structures of the high-fidelity solver, including the discretized operators. Such dependencies often limit its applicability in industrial practice, where commercial solvers (software) are prevalent and source code is inaccessible. Furthermore, for non-linear or non-affine problems, intrusive methods may not provide significant computational gains \cite{romor2023non}. \\

In contrast, non-intrusive approaches are entirely data-driven and do not depend on the governing equations. These methods have demonstrated substantial speed-ups even for non-linear and non-affine problems \cite{Tezzele2022}. Due to their flexibility and minimal implementation constraints, non-intrusive frameworks are particularly attractive in real-world applications. \\

{ In this work, we adopt a non-intrusive reduced-order modelling framework based on Proper Orthogonal Decomposition (POD) for dimensionality reduction, coupled with Radial Basis Function (RBF) interpolation to approximate the mapping from input parameters to the corresponding reduced modal coefficients. This framework enables the construction of an efficient surrogate model for the rapid prediction of residence time distributions in an industrial-scale tundish.} \\

This article extends the previous work on intrusive projection-based ROMs for RTD analysis \cite{Harshith2025PMOR}. In industries, the most common practice is to use commercial solvers for FOMs, which limits access to source code, rendering intrusive ROM development impractical. To overcome this limitation, we develop a non-intrusive, data-driven ROM \cite{Tezzele2022, hesthaven2018non} for RTD analysis, as an alternative compatible with commercial solvers, yielding excellent accuracy across parametric variations, making it well-suited for industrial applications. \\ 

{
The main contributions of this work are as follows:

\begin{itemize}
    \item We develop a purely data-driven, non-intrusive ROM for RTD analysis in an industrial-scale, three-dimensional continuous casting tundish. To the best of the authors’ knowledge, this is the first application of such a ROM for RTD prediction in a full-scale industrial tundish, validated against both experiment and high-fidelity FOM simulations.
    
    \item The ROM enables rapid prediction of RTD responses for multiple unseen parameter instances, offering significant computational savings compared to full-order simulations, and demonstrating strong generalization capabilities within the interpolation regime.
    
    \item The framework provides a practical tool for real-time monitoring, design-space exploration, and process optimization in continuous casting operations, highlighting its potential to support advanced steelmaking decisions. Although a full multi-parameter optimization study is beyond the scope of this work, the proposed ROM lays the foundation for such future analyses.
\end{itemize}}

This paper is organised as follows. Section \ref{sec:FOM} describes the full-order model, including the governing equations, turbulence modelling, Boussinesq approximation, species transport, and the numerical methods employed. In Section \ref{sec:ROM}, the data-driven ROM framework is introduced, comprising the offline and online stages. Section \ref{sec:num_setup} outlines the numerical setup, including the computational domain and boundary conditions. The results and discussion are presented in Section \ref{sec:results}, and finally, conclusions and perspectives are discussed in Section \ref{sec:conclusion}.

%% file: sections/FOM.tex
\section{Full order model}\label{sec:FOM}

In this section, we present the physical and mathematical modelling approach for simulating isothermal and non-isothermal conditions during steady-state tundish operation, where the Reynolds-averaged Navier-Stokes (RANS) equations are employed. Further, we describe the transient tracer transport model employed for the evaluation of RTD curves.

\subsection{Governing equations}

The isothermal, incompressible turbulent flow within the tundish, where the working fluid is water, is governed by the continuity and momentum equations:
\begin{equation}
\nabla \cdot (\rho \mathbf{u}) = 0,
\end{equation}
\begin{equation}
\rho (\mathbf{u} \cdot \nabla) \mathbf{u} = -\nabla P + \nabla \cdot \left[\mu_{\mathrm{eff}} \left( \nabla \mathbf{u} + (\nabla \mathbf{u})^{\top} \right) \right] + \mathbf{F},
\end{equation}
\begin{equation}
\mu_{\mathrm{eff}} = \mu + \mu_t = \mu + \rho c_\mu \frac{k^2}{\varepsilon}.
\end{equation}

Here, $\mathbf{u}$ is the velocity vector, $P$ denotes pressure, and $\mathbf{F}$ represents the body force. The effective viscosity $\mu_{\mathrm{eff}}$ accounts for both the molecular viscosity $\mu$ and the turbulent eddy viscosity $\mu_t$, the latter modelled via the standard $k-\varepsilon$ turbulence framework. The turbulence parameters $k$ and $\varepsilon$ are obtained from the transport equations for turbulent kinetic energy and its dissipation rate, respectively.

\subsection{Boussinesq approximation}

For non-isothermal flow conditions in the tundish, where the working fluid is molten steel, buoyancy effects arising from temperature-dependent density variations are incorporated via the Boussinesq approximation. Under this framework, the momentum equation becomes:

\begin{equation}
\rho (\mathbf{u} \cdot \nabla) \mathbf{u} = -\nabla P + \nabla \cdot \left[ \mu_{\mathrm{eff}} \left( \nabla \mathbf{u} + (\nabla \mathbf{u})^{\top} \right) \right] + \rho_{\text{ref}} \beta \Delta T\, \mathbf{g},
\label{eqn:momentum-Boussinesq}
\end{equation}
where the buoyancy term \( \rho_{\text{ref}} \beta \Delta T\, \mathbf{g} \) accounts for density variation due to temperature difference. Here, \( \rho_{\text{ref}} \) is a reference density, \( \beta \) the thermal expansion coefficient, \( \Delta T = T - T_{\text{ref}} \) the temperature deviation from a reference temperature, and \( \mathbf{g} \) the gravitational acceleration vector. \\

This approximation assumes constant density in all governing equations, except in the buoyancy term of the momentum equation, where linear temperature dependence is retained:
\begin{equation}
(\rho - \rho_{\text{ref}})\mathbf{g} \approx -\rho_{\text{ref}} \beta \Delta T\, \mathbf{g}.
\end{equation}

The Boussinesq approximation is valid when temperature variations are moderate such that \( \beta \Delta T \ll 1 \). For fluids such as water or molten steel, where \( \beta \sim 10^{-4}~\text{K}^{-1} \), this condition holds for temperature differences up to 15–20\,\textdegree C, yielding \( \beta \Delta T \sim 10^{-3} \). \\

The transient temperature field is computed by solving the energy conservation equation:
\begin{equation}
\frac{\partial (\rho C_p T)}{\partial t} + \nabla \cdot (\rho C_p T\, \mathbf{u}) = \nabla \cdot (k_{\mathrm{eff}} \nabla T),
\end{equation}
where \( C_p \) is the specific heat at constant pressure, and \( k_{\mathrm{eff}} \) is the effective thermal conductivity. The temperature field couples with the velocity field through the buoyancy term in equation~\eqref{eqn:momentum-Boussinesq}. In the present simulations, all tundish walls and the free surface are assumed adiabatic, thereby neglecting heat flux through these boundaries.

\subsection{Turbulence model}

Turbulence is modelled using the standard eddy-viscosity based $k$-$\varepsilon$ model \citep{Launder1972}. This approach introduces two additional transport equations: one for the turbulent kinetic energy \( k \), and another for its rate of dissipation \( \varepsilon \), given by
\begin{align}
\rho\, \mathbf{u} \cdot \nabla k &= \nabla \cdot \left( \frac{\mu_{\mathrm{eff}}}{\sigma_k} \nabla k \right) + G_k - \rho \varepsilon, \\
\rho\, \mathbf{u} \cdot \nabla \varepsilon &= \nabla \cdot \left( \frac{\mu_{\mathrm{eff}}}{\sigma_\varepsilon} \nabla \varepsilon \right) + C_1 \frac{\varepsilon}{k} G_k - C_2 \rho \frac{\varepsilon^2}{k},
\end{align}
where \( G_k \) denotes the production of turbulent kinetic energy due to velocity gradients:
\[
G_k = \mu_t \left( \nabla \mathbf{u} + (\nabla \mathbf{u})^{\top} \right) : \nabla \mathbf{u}.
\]

The turbulent viscosity is computed as \( \mu_t = \rho C_\mu \frac{k^2}{\varepsilon} \), and the model constants are: \( C_\mu = 0.09 \), \( C_1 = 1.44 \), \( C_2 = 1.92 \), \( \sigma_k = 1.00 \), and \( \sigma_\varepsilon = 1.30 \). For scalar transport, the turbulent Prandtl number is set to \( \mathrm{Pr}_t = 0.90 \).

\subsection{Species transport equation}

The transient evolution of tracer concentration, represented as a mass fraction \( c_i \), is computed by solving the species transport equation using the previously obtained velocity field:
\begin{equation}
\frac{\partial (\rho c_i)}{\partial t} + \nabla \cdot (\rho \mathbf{u} c_i) = \nabla \cdot \left( \rho D_{\mathrm{eff}} \nabla c_i \right),
\label{eqn:species_transport}
\end{equation}
where \( \rho \) is the fluid density, and \( D_{\mathrm{eff}} \) is the effective diffusivity, given by:
\begin{equation}
D_{\mathrm{eff}} = D_m + D_t = D_m + \frac{\mu_{\mathrm{eff}}}{\mathrm{Pr}_t}.
\end{equation}

Here, \( D_m \) is the molecular diffusivity, \( D_t \) is the turbulent (eddy) diffusivity, \( \mu_{\mathrm{eff}} \) is the effective viscosity defined previously, and \( \mathrm{Pr}_t \) is the turbulent Prandtl number. The transport of species is assumed to be passive, i.e., it does not influence the flow field. \\

The numerical simulations are conducted using the finite-volume library {OpenFOAM} \citep{Weller1998}, an object-oriented C++ framework that supports modular formulation of the governing equations and facilitates solver customization. For steady-state, isothermal flow simulations, the incompressible solver {simpleFoam} is employed and for non-isothermal flow, the {buoyantBoussinesqSimpleFoam} solver is considered, which accounts for buoyancy effects under the Boussinesq approximation. Transient passive tracer transport is simulated using a customized solver, {scalarTurbulentTransportFoam}, which extends the standard {scalarTransportFoam} by incorporating turbulent diffusivity in the scalar transport equation. \\

All flow solvers rely on the SIMPLE algorithm \citep{Caretto} to ensure iterative convergence of the pressure–velocity coupling, and, where applicable, the energy equation. The discretization schemes adopted in the finite-volume formulation are summarized in Table~\ref{table:discretization_scheme}. Second-order accuracy is maintained for the majority of terms, while a first-order upwind scheme is applied to turbulence quantities to enhance numerical stability. For time integration in transient tracer simulations, a second-order implicit backward scheme is employed.

\begin{table}[ht!]
    \centering
    \renewcommand{\arraystretch}{1.2}
    \begin{tabular}{|>{\centering\arraybackslash}p{6cm}|>{\centering\arraybackslash}p{9cm}|}
        %\rowcolor{gray!30}
        \hline
        \textbf{Term} & \textbf{Scheme} \\
        \hline
        Time derivative & steadyState / Second-order implicit backward \\
        Convective term (momentum) & Second-order upwind \\
        Convective term (energy, species) & Second-order upwind \\
        Convective term (turbulence: \(k, \varepsilon\)) & First-order upwind \\
        Diffusive term (viscous, thermal) & Central differing \\
        Gradient term & Limited linear construction \\
        \hline
    \end{tabular}
    \caption{Summary of numerical schemes considered for full-order simulations.}
    \label{table:discretization_scheme}
\end{table}

%% file: sections/ROM.tex
\section{Reduced order model}\label{sec:ROM}

This section presents the data-driven ROM framework for RTD analysis. The framework comprises two stages: offline and online. In the offline stage, a ROM is constructed from high-fidelity simulation data. This involves generating full-order solution snapshots, extracting dominant flow structures using POD, and constructing a low-dimensional linear approximation subspace. The subsequent online phase enables rapid evaluation of the ROM for new parameter instances, with computational efficiency achieved via a regression-based approximation of the parameter-dependent modal coefficients.

\subsection{Offline stage}

{ \textit{Data creation:} To construct the reduced basis space, at first the high-fidelity model introduced in Section~\ref{sec:FOM} is solved for a finite set of parameter values \( \mathcal{K} = \{\boldsymbol{\mu}_k\}_{k=1}^{N_\mu} \subset \mathcal{P} \). Given the time-dependent nature of the problem, the snapshot collection must capture both parametric and temporal variability. Therefore, a discrete set of time instances \( \{t_l\}_{l=1}^{N_t} \subset [0, T] \) is also defined. The total number of snapshots is then \( N_s = N_\mu \cdot N_t \). The snapshot matrix \( \mathcal{S}_k \in \mathbb{R}^{N_h \times N_s} \) is assembled by collecting high-fidelity solutions of the tracer concentration field, evaluated at \( N_h \) spatial degrees of freedom for each parameter-time pair:}
\begin{equation}
    \mathcal{S}_k = \left[ c(\boldsymbol{\mu}_k, t_1), \ldots, c(\boldsymbol{\mu}_k, t_{N_t}) \right] \in \mathbb{R}^{N_h \times N_t}, \quad k = 1, \dots, N_\mu.
\end{equation}
\subsubsection{  Proper Orthogonal Decomposition}
To generate a low-dimensional subspace suitable for projection-based model reduction, several techniques are available in the literature~\cite{quarteroni2015reduced, hesthaven2016certified, rozza2022advanced}, including Proper Orthogonal Decomposition (POD), Proper Generalized Decomposition (PGD), and reduced basis (RB) methods based on greedy sampling. Among model order reduction techniques, POD is widely adopted for its ability to extract the most energetic modes from computational fluid dynamics (CFD) simulations \cite{rozza2022advanced}.\\

Let \( c(\mu, t) \) denote a scalar or vector-valued field, with a corresponding collection of \( N_s \) snapshot realizations \( \{c_i\}_{i=1}^{N_s} \). The goal of POD is to identify, for each desired reduced dimension \( N_{\text{POD}} = 1, \ldots, N_s \), an optimal set of orthonormal basis functions \( \{\varphi_j\}_{j=1}^{N_{\text{POD}}} \) that best approximates the snapshots in the least-squares sense. Specifically, the POD basis minimizes the total projection error:

\begin{equation}
\min_{\{\varphi_j\}_{j=1}^{N_{\text{POD}}}} \sum_{i=1}^{N_s} \left\| c_i - \sum_{j=1}^{N_{\text{POD}}} \langle c_i, \varphi_j \rangle_{L^2(\Omega)} \varphi_j \right\|^2_{L^2(\Omega)}.
\label{eqn:POD_minimization}
\end{equation}

These basis functions must also satisfy the orthonormality condition:
\begin{equation}
\langle \varphi_i, \varphi_j \rangle_{L^2(\Omega)} = \delta_{ij}, \quad \forall i, j = 1, \ldots, N_{\text{POD}}.
\end{equation}

As shown in~\cite{quarteroni2015reduced}, the minimization problem in Equation~\eqref{eqn:POD_minimization} is equivalent to solving the eigenvalue problem associated with the correlation matrix \( \boldsymbol{\mathcal{C}} \in \mathbb{R}^{N_s \times N_s} \), defined as:

\begin{equation}
    \boldsymbol{\mathcal{C}} \boldsymbol{Q} = \boldsymbol{Q} \boldsymbol{\Lambda},
\end{equation}
\begin{equation}
    \boldsymbol{\mathcal{C}}_{ij} = \langle c_i, c_j\rangle_{L^2(\Omega)}, \quad \forall i, j = 1, \ldots, N_s.
\end{equation}
where \( \boldsymbol{Q} \) contains the eigenvectors and \( \boldsymbol{\Lambda} \) is the diagonal matrix of eigenvalues \( \{\lambda_i\}_{i=1}^{N_s} \). The POD basis functions are then reconstructed as a linear combination of the original snapshots:
\begin{equation}
\varphi_i = \frac{1}{\sqrt{\lambda_i}} \sum_{j=1}^{N_s} c_j (\boldsymbol{Q})_{ij}.
\end{equation}

The POD basis functions obtained through the methodology described above define a reduced approximation subspace given by:
\begin{equation}
    V_{\mathrm{rb}} = \text{span} \left\{ \varphi_i \right\}_{i=1}^{N_{\mathrm{rb}}} \subset L^2(\Omega),
\end{equation}
where \( N_{\mathrm{rb}} \ll N_h \) is the dimension of the reduced basis space, typically determined by analyzing the decay of the associated POD eigenvalues. \\

Once the reduced subspace is identified, the high-fidelity concentration field \( c(\mu, t) \) can be approximated by a linear combination of the POD modes:
\begin{equation}
    c(\mu, t) \approx \sum_{i=1}^{N_{\mathrm{rb}}} a_i(\mu, t) \, \varphi_i(x),
    \label{eqn:linearapproximation}
\end{equation}
where \( \{a_i(\mu, t)\}_{i=1}^{N_{\mathrm{rb}}} \) denote the parameter-time dependent modal coefficients, and \( \{\varphi_i(x)\}_{i=1}^{N_{\mathrm{rb}}} \) are the orthonormal, parameter-time independent POD basis functions. \\

The reduced basis matrix \( V_{\mathrm{rb}} \in \mathbb{R}^{N_h \times N_{\mathrm{rb}}} \) is formed by arranging the POD modes as column vectors. The corresponding reduced coefficients \( a(\mu, t_i) \in \mathbb{R}^{N_{\mathrm{rb}}} \) are computed by projecting the full-order solution \( c(\mu, t_i) \in \mathbb{R}^{N_h} \) onto the reduced basis space:
\begin{equation}
    a(\mu, t_i) = V_{\mathrm{rb}}^\top\, c(\mu, t_i).
\end{equation}

The POD basis is computed using the method of snapshots \cite{Sirovich1987}. {In this work, we apply a nested POD procedure, performing a first-stage POD in the temporal domain for each parameter, followed by a second-stage POD across the parametric dimension, as proposed in~\cite{stabile2018finite, rozza2022advanced}. The basis generation is performed using ITHACA-FV \footnote{The ITHACA-FV source code is available at \texttt{https://github.com/ITHACA-FV/ITHACA-FV}.} \cite{ITHACA-FV}, an open-source OpenFOAM based library for reduced order modelling of parametrized problems.}

\subsubsection{  Solution manifold approximation}\label{subsec:solmanifold}
To enable efficient prediction of solution snapshots for unseen parameter instances, it is necessary to construct a mapping from the parameter space to the space of reduced coefficients. This task involves approximating the solution manifold, represented in terms of parameter-time dependent modal coefficients associated with the reduced basis, using data from high-fidelity simulations. \\

Given a set of \( N_s = N_\mu \cdot N_t \) input-output pairs \( \{({\mu}_k, t_l), {a}_{kl}\}_{k=1, l=1}^{N_\mu, N_t} \), where \( {a}_{kl} \in \mathbb{R}^{N_{\mathrm{rb}}} \) are the reduced coefficients computed via projection at time \( t_l \) and parameter \( \boldsymbol{\mu}_k \), the goal is to approximate the mapping:
\begin{equation}
f: (\boldsymbol{\mu}, t) \in \mathbb{ R \times R}^{d_\mu} \mapsto {a}(\boldsymbol{\mu}, t) \in \mathbb{R}^{N_{\mathrm{rb}}}.
\label{eqn:mapping_param_to_coeff}
\end{equation}

Several techniques are available for learning the map \( f \), including linear interpolation, radial basis function (RBF) interpolation~\cite{Xiao2015, taddei2020registration}, Gaussian process regression (GPR) \cite{Tezzele2022GPR}, inverse distance weighting (IDW), multi-fidelity methods, and artificial neural networks (ANNs)~\cite{hesthaven2018non, Tezzele2022}. {{In this study, RBF interpolation is employed to approximate the mapping from parameter-time pairs $(\boldsymbol{\mu}, t)$ to the POD modal coefficients $a_j$, $j=1,\dots,N_\mathrm{rb}$. For a new input $(\boldsymbol{\mu}^*, t^*)$, the $j$-th coefficient is predicted as
\begin{equation}
f_j(\boldsymbol{\mu}^*, t^*) = \sum_{k=1}^{N_\mu} \sum_{l=1}^{N_t} w_{kl}^{(j)} \, \phi\Big( \| (\boldsymbol{\mu}^*, t^*) - (\boldsymbol{\mu}_k, t_l) \|_2 \Big),
\end{equation}
where $\{(\boldsymbol{\mu}_k, t_l)\}$ are the training samples, $w_{kl}^{(j)}$ are weights determined by enforcing interpolation of the known training modal coefficients $a_{kl}^{(j)}$, and $\phi(\cdot)$ is a radial basis function. In this work, a thin-plate spline kernel is employed, defined as $\phi(r) = r^2 \, \log(r)$. The RBF interpolation is implemented using the \texttt{SciPy} library in Python. RBF interpolation is employed due to its simplicity and effectiveness in low-dimensional parametric settings, as considered in the present study. However, for higher-dimensional parametric settings, alternative regression techniques such as Gaussian Process Regression \cite{Tezzele2022GPR} or neural networks \cite{hesthaven2018non} may be more suitable.}

\subsection{  Online stage}

Once the mapping is constructed, the reduced order solution for any unseen input pair \( (\boldsymbol{\mu}^*, t^*) \) is efficiently computed as:
\begin{equation}
\tilde{c}(\boldsymbol{\mu}^*, t^*) = \sum_{j=1}^{N_{\mathrm{rb}}} f_j(\boldsymbol{\mu}^*, t^*) \, \varphi_j(x),
\end{equation}
where \( f_j(\boldsymbol{\mu}^*, t^*) \) is the predicted \( j \)-th modal coefficient, and \( \{\varphi_j(x)\}_{j=1}^{N_{\mathrm{rb}}} \) are the reduced basis functions. This enables accurate, low-cost approximations of the concentration field at new parameter instances.} 

{

\subsection{Data-driven reduced order model algorithm}

The proposed data-driven ROM consists of an offline and an online stage, as summarized in Algorithm~\ref{Algorithm1}. In the offline stage, high-fidelity solutions \( c(\boldsymbol{\mu}_k, t_l) \) are computed over a set of training parameters and time instances and assembled into individual snapshot matrices capturing temporal dynamics for each parameter. A nested POD is then applied: a temporal POD is performed for each parameter instance, followed by a parametric POD on the resulting temporal modes to extract a reduced basis \( \{\varphi_i\}_{i=1}^{N_{\mathrm{rb}}} \), defining a low-dimensional approximation space. The corresponding modal coefficients are obtained by projecting the high-fidelity data onto this reduced space. A regression model is subsequently trained to approximate the mapping \( f: (t, \boldsymbol{\mu}) \mapsto a(t, \boldsymbol{\mu}) \), providing a data-driven representation of the solution manifold. In the online stage, for any new input pair \( (\boldsymbol{\mu}^*, t^*) \), the predicted coefficients are used to reconstruct the approximate solution as a linear combination of the reduced basis functions. \\

\begin{algorithm}
\caption{Data-driven Reduced Order Model}
\label{Algorithm1}
\begin{algorithmic}[1]
\vspace{2mm}

\STATE \textbf{Offline Stage:}

\STATE Define parameter training set 
\( \mathcal{K} = \{\boldsymbol{\mu}_k\}_{k=1}^{N_\mu} \) and time instances 
\( \{t_l\}_{l=1}^{N_t} \);

\STATE Compute high-fidelity snapshots:
\(
c(\boldsymbol{\mu}_k, t_l), \quad \forall k = 1,\dots,N_\mu,\; l = 1,\dots,N_t;
\)

\STATE Assemble snapshot matrix 
\(
\mathcal{S} = \left[ c(\boldsymbol{\mu}_k, t_l) \right] \in \mathbb{R}^{N_h \times N_s}, 
\quad N_s = N_\mu N_t;
\)

\STATE Perform nested POD:
\begin{itemize}
    \item Temporal POD for each parameter instance;
    \item Parametric POD on temporal modes;
\end{itemize}

\STATE Extract reduced basis functions 
\( \{\varphi_i\}_{i=1}^{N_{\mathrm{rb}}} \) and construct reduced space
\(
V_{\mathrm{rb}} = \text{span}\{\varphi_i\}_{i=1}^{N_{\mathrm{rb}}};
\)

\STATE Compute modal coefficients via projection:
\(
a(\boldsymbol{\mu}_k, t_l) = V_{\mathrm{rb}}^\top \, c(\boldsymbol{\mu}_k, t_l);
\)

\STATE Solution manifold approximation to the learn mapping:
\[
f: (t, \boldsymbol{\mu}) \mapsto a(t, \boldsymbol{\mu});
\]

\STATE \textbf{Online Stage:}

\STATE Given new input \( (\boldsymbol{\mu}^*, t^*) \), evaluate:
\(
\hat{a}(t^*, \boldsymbol{\mu}^*) = f(t^*, \boldsymbol{\mu}^*);
\)

\STATE Compute reduced-order solution:
\(
\tilde{c}(\boldsymbol{\mu}^*, t^*) 
= \sum_{i=1}^{N_{\mathrm{rb}}} \hat{a}_i(t^*, \boldsymbol{\mu}^*) \, \varphi_i;
\)

\end{algorithmic}
\end{algorithm}

To assess the accuracy of the ROM predictions, we evaluate the error on the quantity of interest (QoI), defined as the tracer concentration measured at the tundish outlet, i.e., the RTD curve. The relative $L_2$ error between the ROM prediction \( \tilde{c}(t) \) and the high-fidelity solution \( c(t) \) is given by
\begin{equation}
\mathcal{L}_{2} = 
\frac{\| c - \tilde{c} \|_2}{\| c \|_2} 
= \frac{\sqrt{\sum_{i=1}^{N_t} \left( c( \boldsymbol{\mu}, t_i) - \tilde{c}(\boldsymbol{\mu}, t_i) \right)^2}}{\sqrt{\sum_{i=1}^{N_t} c(\boldsymbol{\mu}, t_i)^2}} \,,
\label{eqn:l2_error}
\end{equation}
where \( N_t \) is the number of temporal samples, for a given parameter \( \boldsymbol{\mu}\).

}

%% file: sections/domain.tex
\section{Numerical setup}\label{sec:num_setup}

This section outlines the numerical framework employed for full-order model simulations. The computational geometry and mesh configuration of the tundish system are presented in Section~\ref{subsec:computational_domain}, while the boundary conditions for both isothermal and non-isothermal cases are detailed in Section~\ref{BCs}.

\subsection{Computational domain}\label{subsec:computational_domain}

\begin{figure}[ht!]
    \centering
    \includegraphics[width=0.75\linewidth]{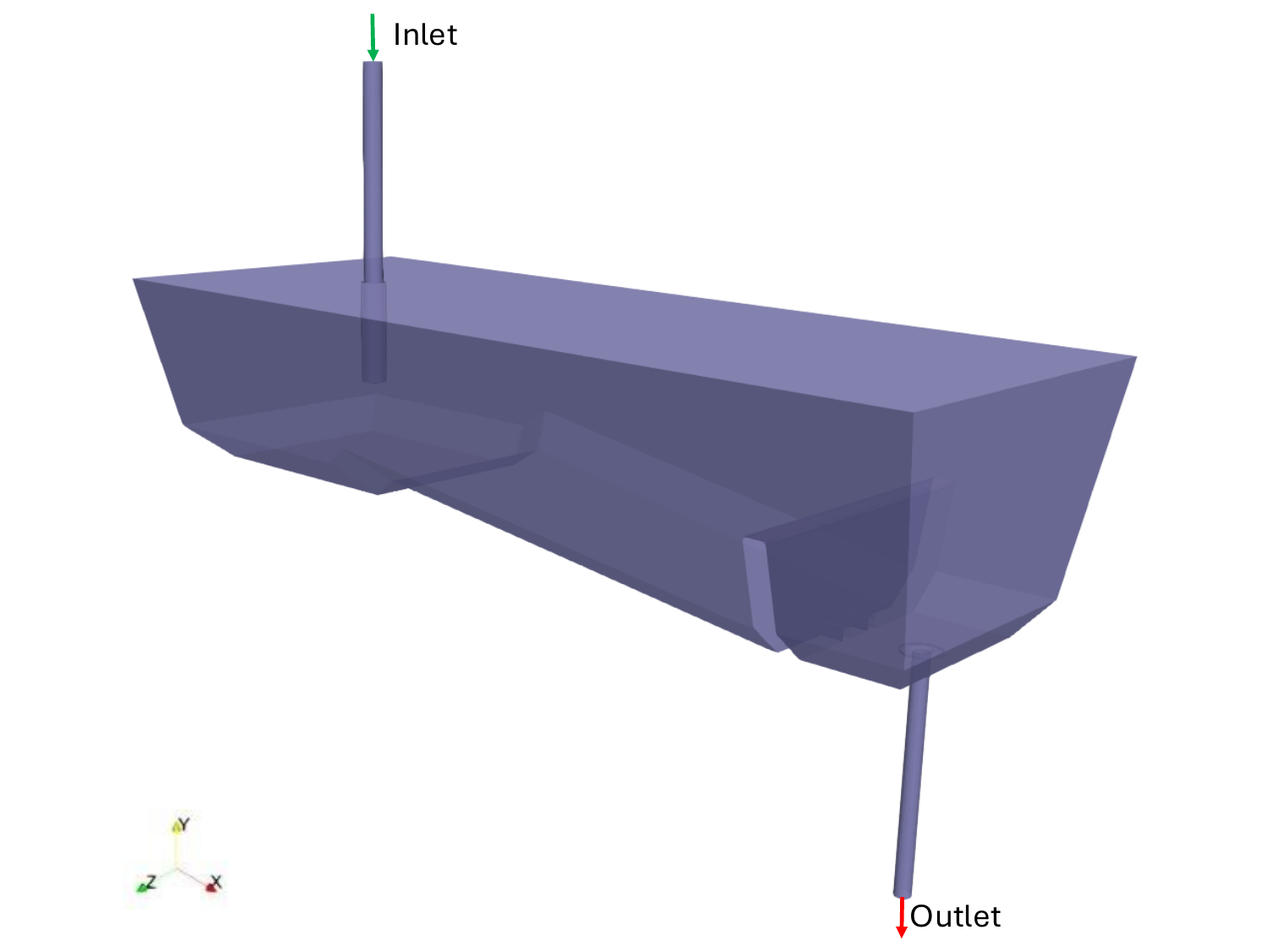}
    \caption{Isometric view of the computational domain for the 1:1 scaled single-strand tundish. The inlet and outlet locations are indicated; all remaining boundaries are considered as no-slip walls.}
    \label{fig:geometry}
\end{figure}

\begin{figure}[ht!]
    \centering
    \includegraphics[width=1\linewidth]{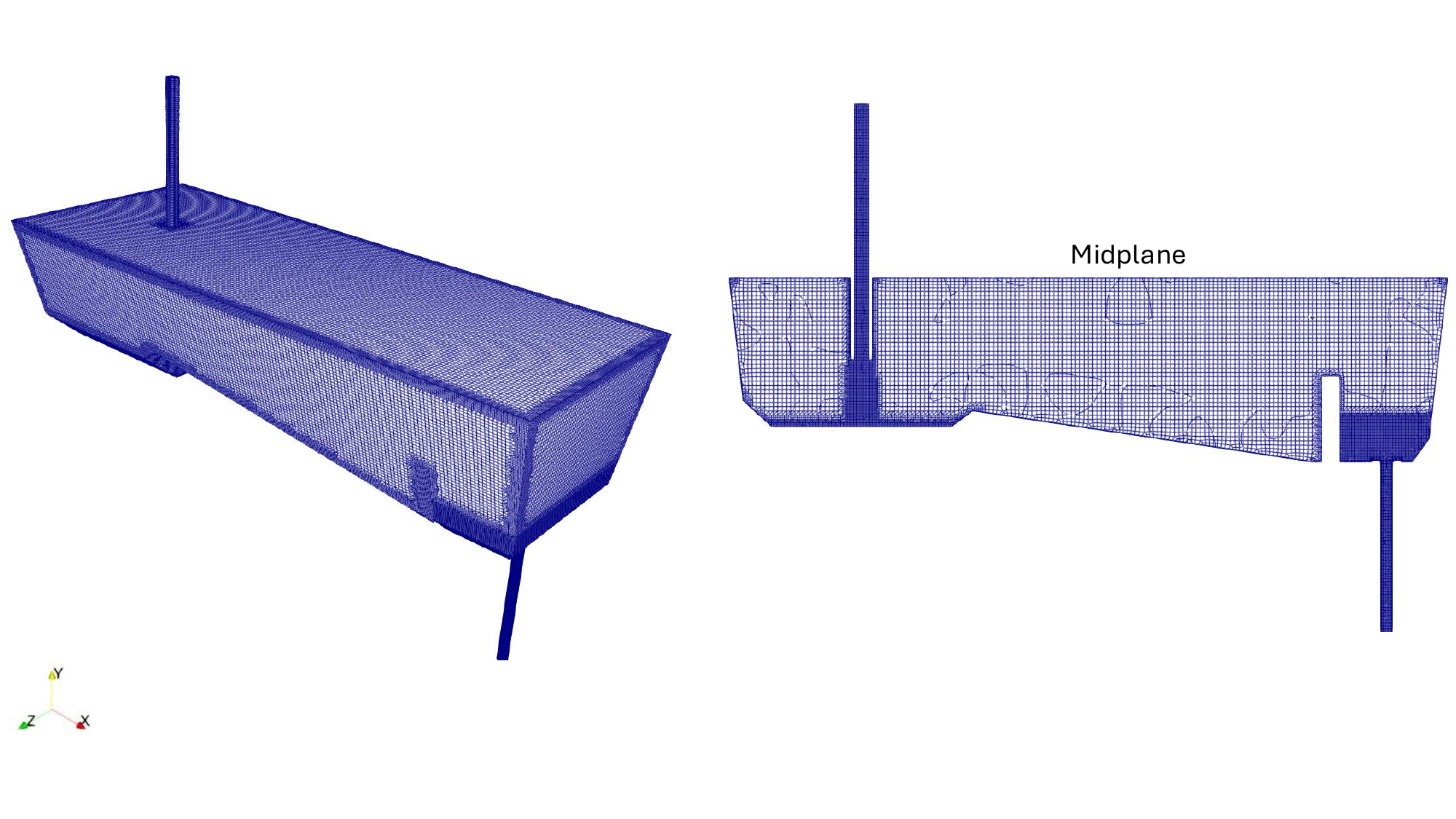}
    \caption{Discretized computational domain of a 1:1 scaled single-strand tundish. \textbf{Left}: Isometric view of the mesh; \textbf{Right}: 2D midplane slice showing mesh refinement near the inlet jet impingement region and outlet, and with six prism layers adjacent to the walls.}
    \label{fig:mesh}
\end{figure}

A three-dimensional computational domain is constructed based on a full-scale (\(1{:}1\)) industrial tundish, as illustrated in Figure~\ref{fig:geometry}, indicating inlet, outlet, and wall boundaries. The mesh was generated using \text{snappyHexMesh} in OpenFOAM, resulting in approximately 3.6 million cells. The mesh is predominantly composed of hexahedra (97\%), with additional prism layers, pyramids, tetrahedra, and polyhedra near geometrically complex regions (Figure~\ref{fig:mesh}).


\subsection{Boundary Conditions}\label{BCs}

A constant volumetric flow rate is prescribed at the inlet, outflow conditions are applied at the outlet, and no-slip conditions are enforced on all walls using standard wall functions. For non-isothermal simulations, lateral and bottom wall heat losses are set to \(6~\mathrm{kW}\,\mathrm{m}^{-2}\). A summary of input parameters and boundary conditions is provided in Table~\ref{table:inputparameters}. \\

Two simulation scenarios are considered. First, an isothermal water case, leverages dynamic similarity (matched Froude number) to replicate molten steel flow characteristics and to validate the model against water experiments. Second, a non-isothermal molten steel case, incorporates buoyancy effects via the Boussinesq approximation to capture thermal stratification relevant to industrial tundish operation. Together, these cases enable both model validation and demonstration under realistic continuous casting conditions. \\

\begin{table}[ht!]
\centering
\begin{tabular}{|c|c|c|}
\hline
%\rowcolor{gray!30}
 & \textbf{Water} & \textbf{Molten steel} \\
%\rowcolor{gray!30}
\textbf{Parameter} & \textbf{(Isothermal)} & \textbf{(Non-isothermal)} \\
\hline
Density & $1000~\mathrm{kg}/\mathrm{m}^3$ & $6900~\mathrm{kg}/\mathrm{m}^3$ \\
Kinematic viscosity & $1 \times 10^{-6}~\mathrm{m^2}/\mathrm{s}$ & $0.8 \times 10^{-6}~\mathrm{m^2}/\mathrm{s}$ \\
Reference pressure & $101{,}325~\mathrm{Pa}$ & $101{,}325~\mathrm{Pa}$ \\
Heat capacity & -- & $800~\mathrm{J}/\mathrm{kg}\cdot\mathrm{K}$ \\
Thermal conductivity & -- & $35~\mathrm{W}/\mathrm{m}\cdot\mathrm{K}$ \\
Thermal expansion coefficient & -- & $0.000127~\mathrm{K}^{-1}$ \\
Inlet volumetric flow rate  & $0.0102657~\mathrm{m^3}/\mathrm{s}$ & $0.0102657~\mathrm{m^3}/\mathrm{s}$ \\
Inlet velocity & $1.6~\mathrm{m}/\mathrm{s}$ & $1.6~\mathrm{m}/\mathrm{s}$  \\
Inlet temperature & -- & $1823.15~\mathrm{K}$ \\
Wall (flow) & No-slip & No-slip \\
Wall (heat loss) & -- & $6~\mathrm{kW}/\mathrm{m}^2$ \\
\hline
\end{tabular}
\caption{ Summary of input parameters and boundary conditions used in the isothermal and non-isothermal simulations.}
\label{table:inputparameters}
\end{table}

\begin{figure}[ht!]
    \centering
    \includegraphics[width=0.65\linewidth]{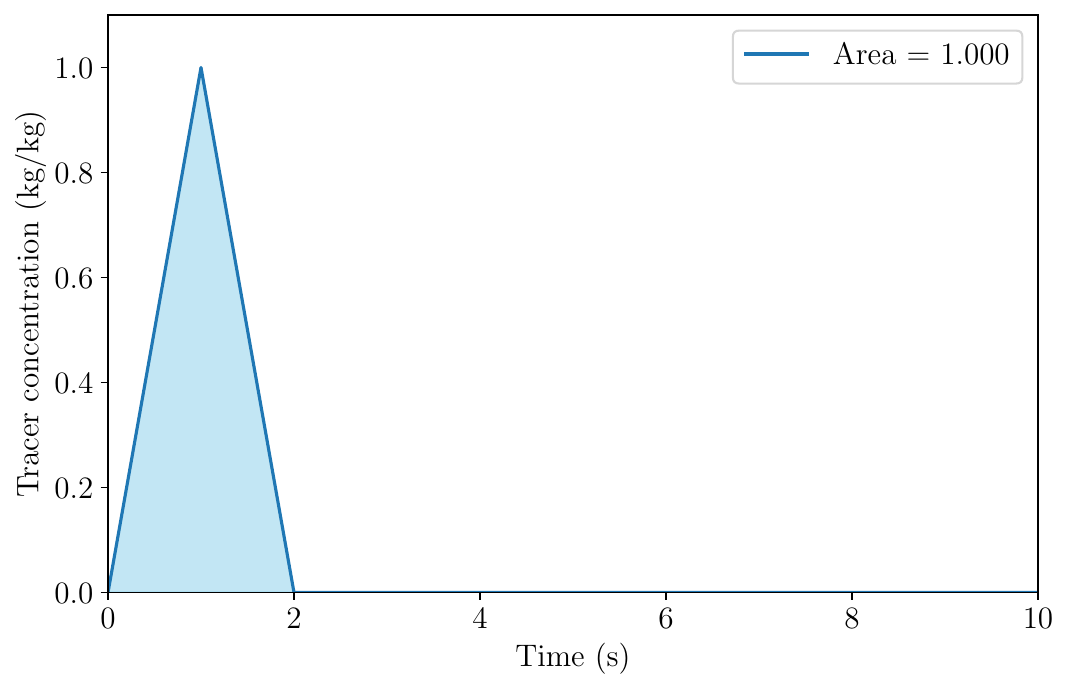}
    \caption{Time-dependent tracer concentration profile imposed at the inlet for the transient species transport simulation.}
    \label{fig:Tracer_inletBC}
\end{figure}

For tracer injection, a pulse input is applied by setting the tracer concentration (mass fraction) at the inlet to unity for \(0 \leq t \leq 2~\mathrm{s}\), and zero thereafter. The outlet concentration is recorded over \(0 \leq t \leq 3000~\mathrm{s}\) to generate the RTD curves (see Figure~\ref{fig:Tracer_inletBC}).

%% file: sections/numerical_experiments.tex
\section{Results and discussion}\label{sec:results}
\begin{figure}
    \centering
    \includegraphics[width=0.75\linewidth]{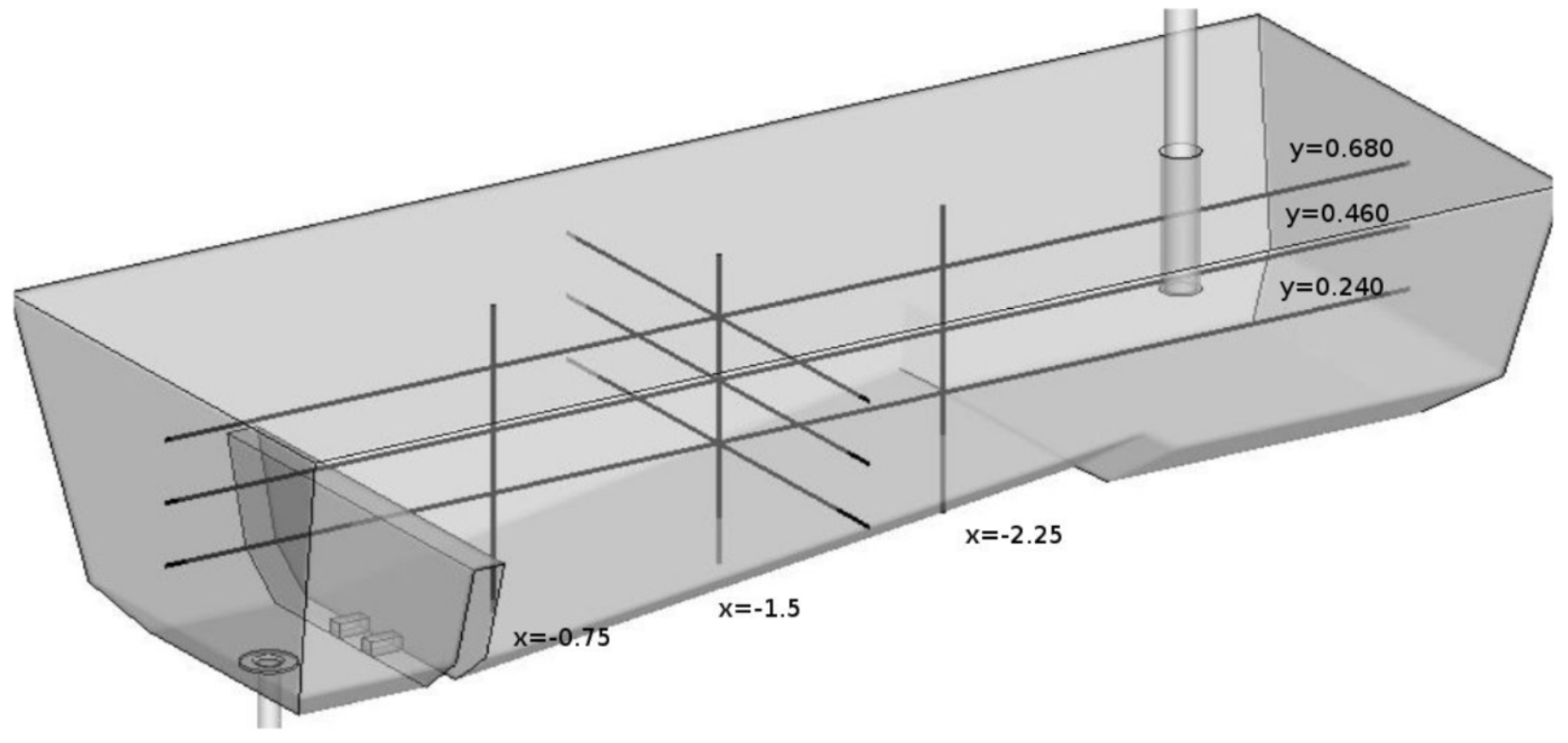}
    \caption{Positions of the horizontal and vertical lines along which streamwise velocity profiles are extracted within the tundish.}
    \label{fig:tundish_Vlines}
\end{figure}

\begin{figure}[ht!]
    \centering
    \includegraphics[width=1\linewidth]{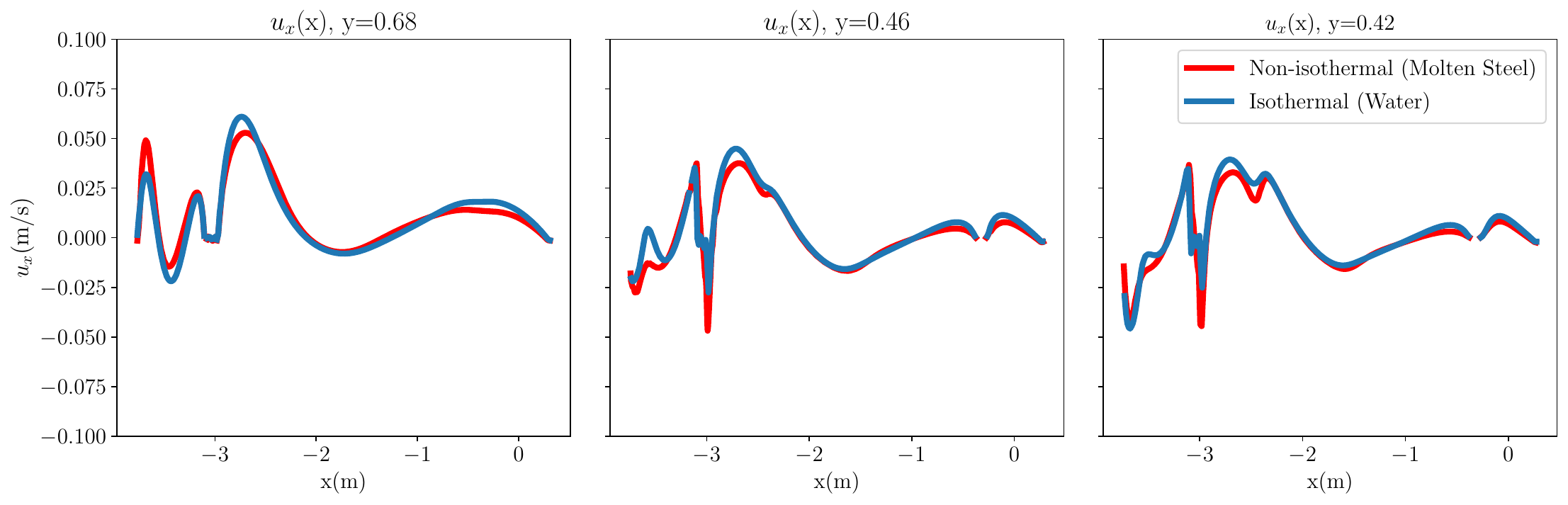}
    \caption{Distributions of streamwise velocity along horizontal lines at selected y-positions (illustrated in Figure \ref{fig:tundish_Vlines}), for both isothermal and non-isothermal cases.}
    \label{fig:velocity-horizontal-line}
\end{figure}

\begin{figure}[ht!]
    \centering
    \includegraphics[width=1\linewidth]{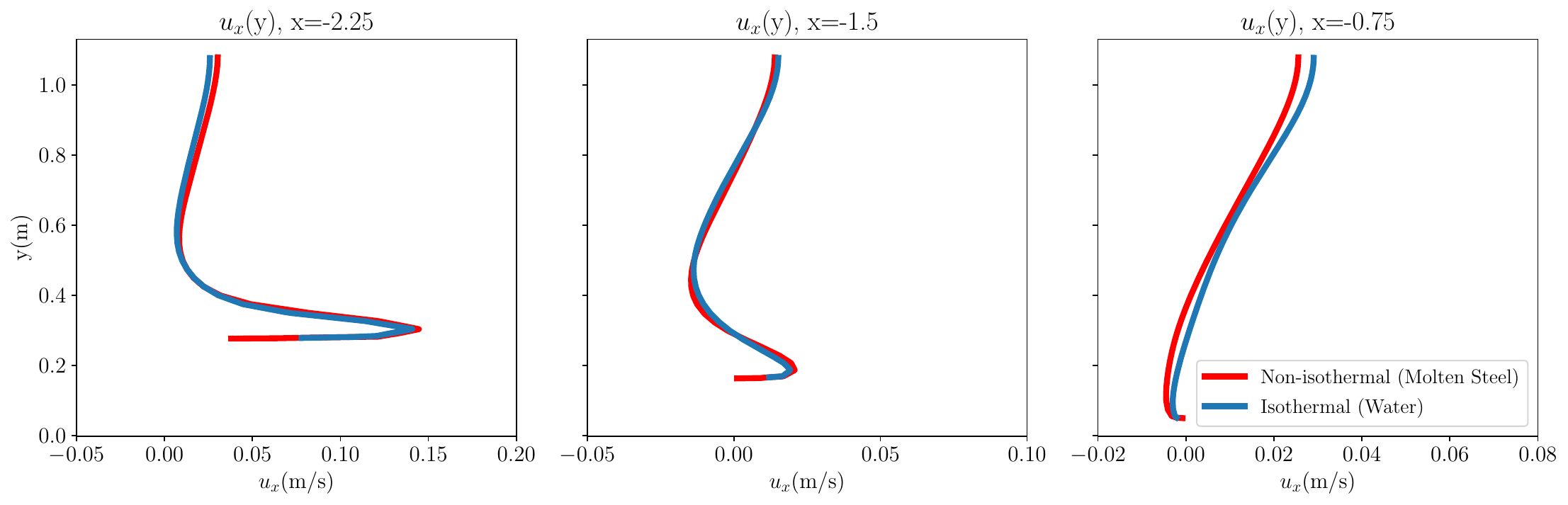}
    \caption{Distributions of streamwise velocity measured along vertical lines at selected x-positions (illustrated in Figure \ref{fig:tundish_Vlines}), for both isothermal and non-isothermal cases.}
    \label{fig:velocity-vertical-line}
\end{figure}

In this study, three-dimensional steady-state simulations of turbulent flow are conducted under both isothermal (water) and non-isothermal (molten steel) conditions. To evaluate the spatial variation of the streamwise velocity component \(u_x\), velocity profiles are extracted along selected horizontal and vertical lines within the tundish, as illustrated in Figure~\ref{fig:tundish_Vlines}. Figure~\ref{fig:velocity-horizontal-line} compares the streamwise velocity distributions at various lateral (\(y\)) positions for the two thermal conditions, while Figure~\ref{fig:velocity-vertical-line} presents corresponding profiles along vertical sections at different streamwise (\(x\)) locations. \\


The velocity distributions under isothermal and non-isothermal conditions show close agreement, indicating that buoyancy effects have minimal influence on the flow field. These distributions, along with tracer studies, form the foundation for the subsequent RTD analysis and volume partitioning, which are used to assess the mixing efficiency and flow characteristics within the tundish. \\

\begin{figure}[ht!]
    \centering
    \includegraphics[width=1\linewidth]{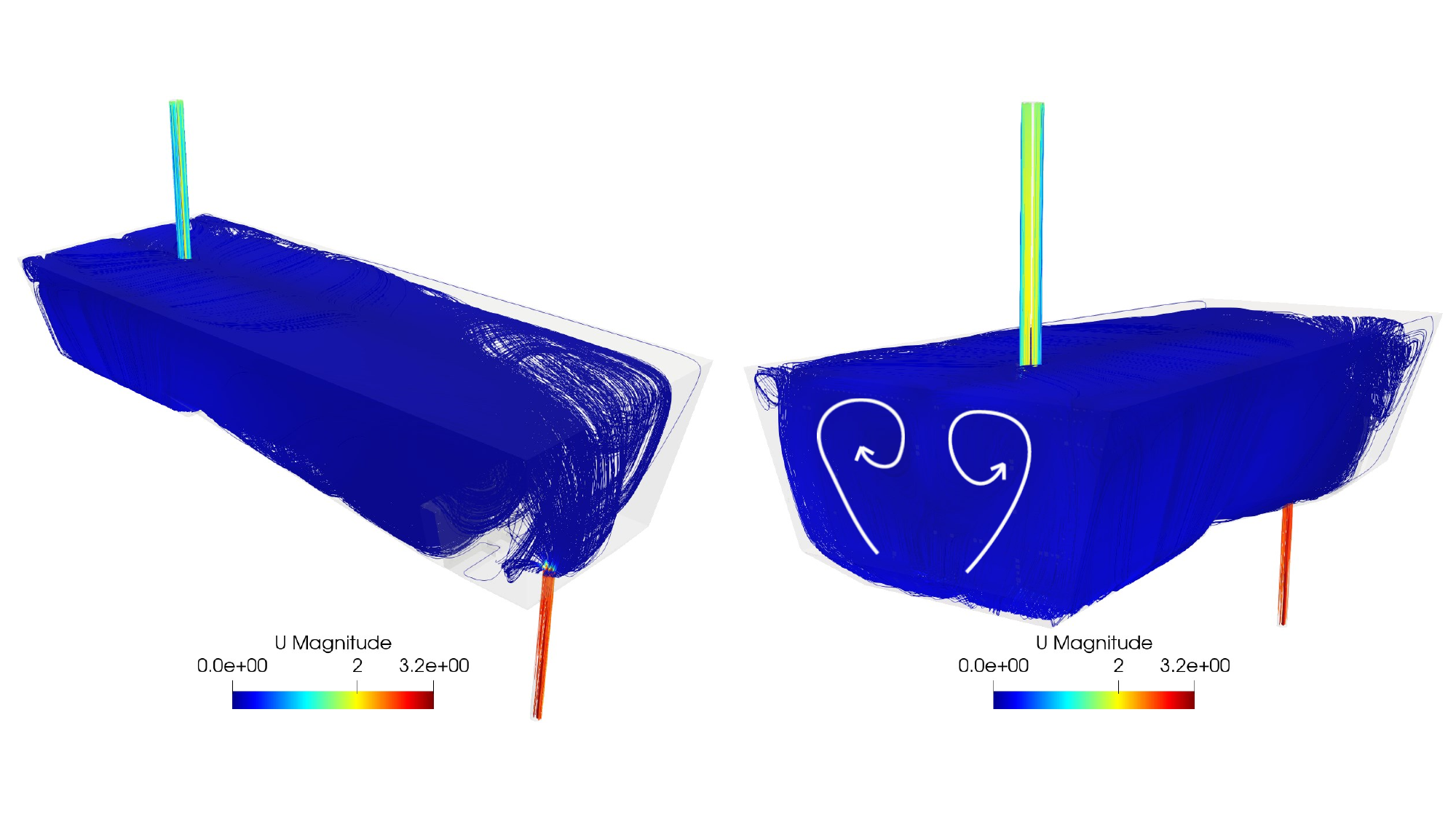}
    \caption{Streamline paths visualized in the tundish using velocity magnitude with inlet velocity 1.6 \si{m/s}. The impingement of the inlet jet induces a circulation zone, which contributes to enhanced mixing characteristics, increased residence time, and mitigated short-circuiting effects. }
    \label{fig:steamline}
\end{figure}

{{{As shown in Figure~\ref{fig:steamline}, the streamline paths in the inlet jet impingement region indicate the formation of a circulation zone within the tundish, which governs the internal flow structure. This recirculating region enhances mixing by redistributing momentum and reducing direct transport from inlet to outlet. Consequently, it increases fluid residence time and mitigates short-circuiting effects.}}}\\

Transient tracer injection simulations are performed to characterize the flow behavior. The resulting tracer concentration–time profile, known as the RTD curve, serves as the basis for the RTD-based analysis. Figure~\ref{fig:RTD-analysis} shows the RTD curve derived from the CFD simulations, highlighting key metrics such as \(T_{\min}, T_{\max}, T_{\text{avg}},\) and \(T_{\text{th}}\). The corresponding flow partitioning into plug volume (\(V_{\text{plug}}\)), mixing volume (\(V_{\text{mix}}\)), and dead volume (\(V_{\text{dead}}\)) is illustrated in Figure~\ref{fig:volume_partition}, which is widely used to quantify the influence of internal flow structures on stagnation and active transport regions. \\

{{{RTD curves are significantly influenced by tundish flow modifiers such as impact pads, dams, and ceramic filters, which modify the internal velocity field and circulation structures, thereby affecting residence time distribution and dead volume \cite{Tkadlekov2020, Bensouici2009}. In addition, RTD behavior is highly sensitive to flow rate variations, highlighting the dependence of tundish performance on operating conditions, such that a design configuration that performs well under one regime may become suboptimal under another \cite{anapagaddi2013exploring, Anapagaddi2014}.}} \\

\begin{figure}[ht!]
    \centering
    \includegraphics[width=0.7\linewidth]{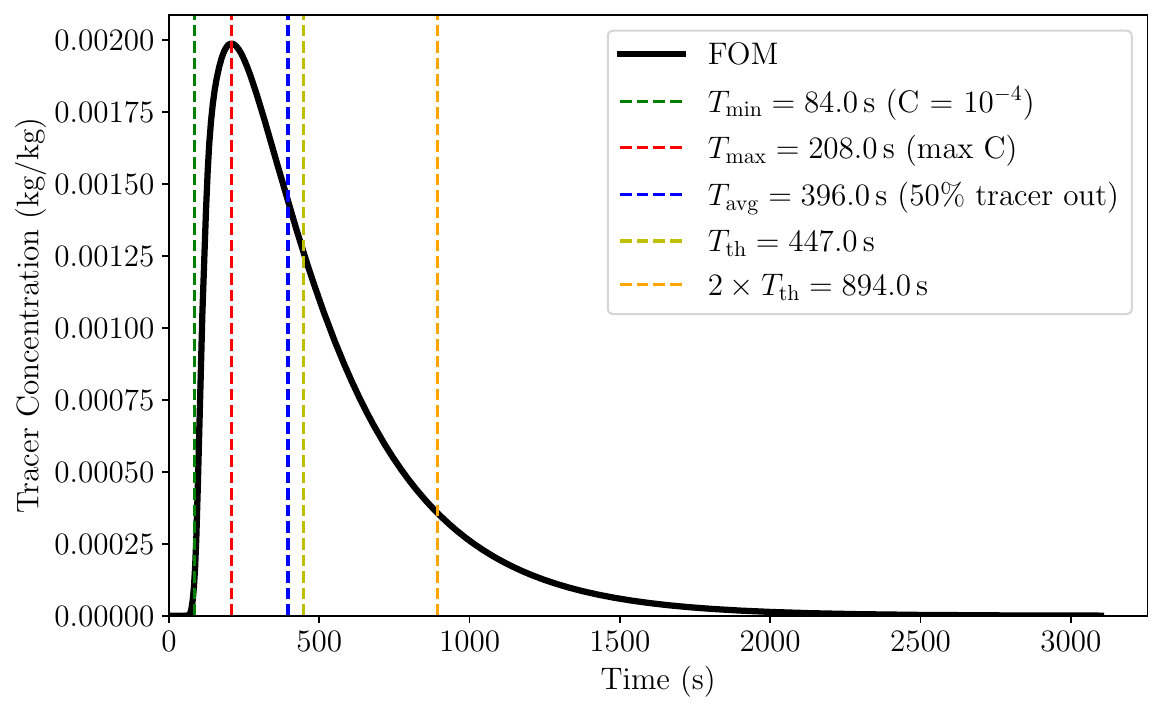}
    \caption{RTD curve derived from FOM simulations, illustrating key flow characteristics required for volume partitioning.}
    \label{fig:RTD-analysis}
\end{figure}
\begin{figure}[ht!]
    \centering
    \includegraphics[width=0.6\linewidth]{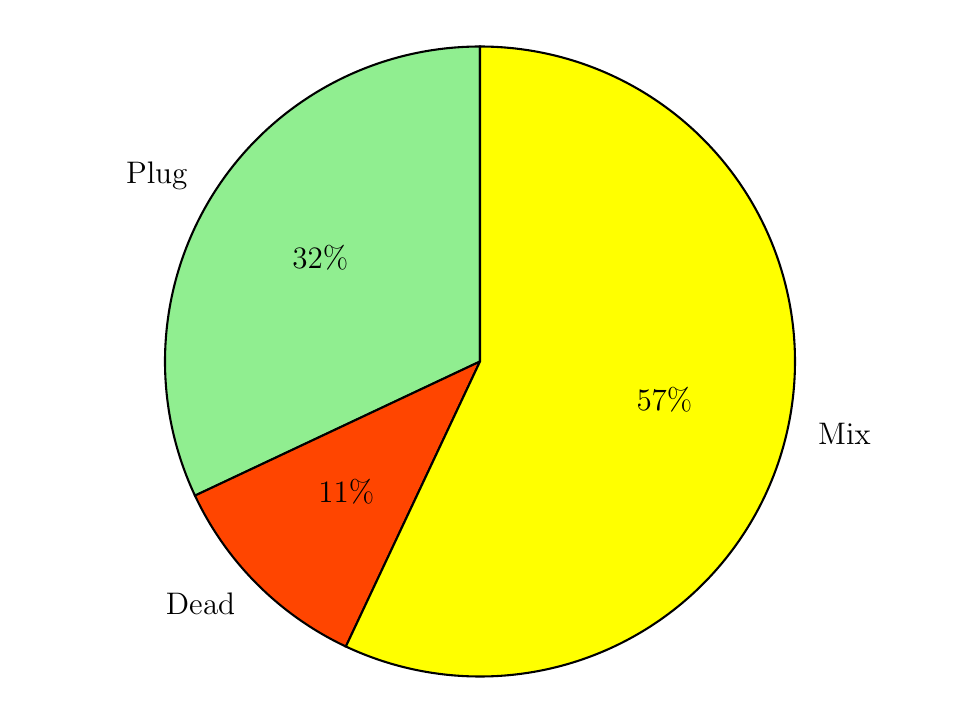}
    \caption{Volume partitioning obtained from RTD analysis of a FOM simulation result.}
    \label{fig:volume_partition}
\end{figure}

{In industrial settings, high-fidelity full-order simulations are computationally expensive, making extensive parametric studies prohibitive and limiting the availability of training data for ROM construction. Therefore, a design of experiments (DoE) is constructed by varying the inlet velocity (or equivalently, the throughput rate), which is a dominant parameter influencing the RTD behaviour in tundish flows. \\

Table~\ref{Table:RTD_parameters} summarizes the inlet velocities and corresponding flow rates considered for the parametric study of RTD behaviour in a single-strand tundish. These cases are used to assess the sensitivity of RTD characteristics to variations in flow conditions. While the present study focuses on a one-dimensional parameter space, this choice reflects practical industrial constraints and serves as a first step toward more comprehensive multi-parameter analyses. \\

For each inlet velocity listed in Table~\ref{Table:RTD_parameters}, an isothermal steady-state simulation is first conducted to compute the velocity field. Subsequently, transient scalar transport simulations are performed to obtain the RTD curves, shown in Figure~\ref{fig:RTD_FOM_results}. As the inlet velocity decreases, the RTD curves broaden, the peak concentration diminishes, and the profiles become increasingly dispersed. This trend reflects the longer residence times and reduced mixing efficiency associated with slower flow rates. The results emphasize the strong dependence of residence time distribution on inlet velocity, with broader RTD curves indicating diminished flow uniformity and reduced system performance. The resulting dataset forms the basis for training and evaluating the proposed ROM, which enables rapid prediction across parameter instances that would otherwise require computationally expensive FOM simulations. \\

\begin{table}[ht!]
    \centering
    \renewcommand{\arraystretch}{1}
    \begin{tabular}{|>{\centering\arraybackslash}p{3cm}|
                    >{\centering\arraybackslash}p{4cm}|
                    >{\centering\arraybackslash}p{4cm}|}
        \hline
        \textbf{Parameter No.} & \textbf{Flow rate (\si{m^3/hr})} & \textbf{Velocity (\si{m/s})} \\
        \hline
        1 & 36.9365 & 1.6301 \\
        2 & 34.52985 & 1.5239 \\
        3 & 32.1232 & 1.4177 \\
        4 & 29.71655 & 1.3115 \\
        5 & 27.3099 & 1.2053 \\
        6 & 24.90325 & 1.0991 \\
        7 & 22.4966 & 0.9929 \\
        8 & 20.08995 & 0.8866 \\
        9 & 17.6833 & 0.7804 \\
        10 & 15.27665 & 0.6742 \\
        11 & 12.87 & 0.5680 \\
        \hline
    \end{tabular}
    \caption{Inlet velocities and flow rates considered in the parametric study of RTD curves for the single-strand tundish. These parameters were varied to conduct a sensitivity analysis of the RTD characteristics.}
    \label{Table:RTD_parameters}
\end{table}

\begin{figure}[ht!]
    \centering
    \includegraphics[width=0.8\linewidth]{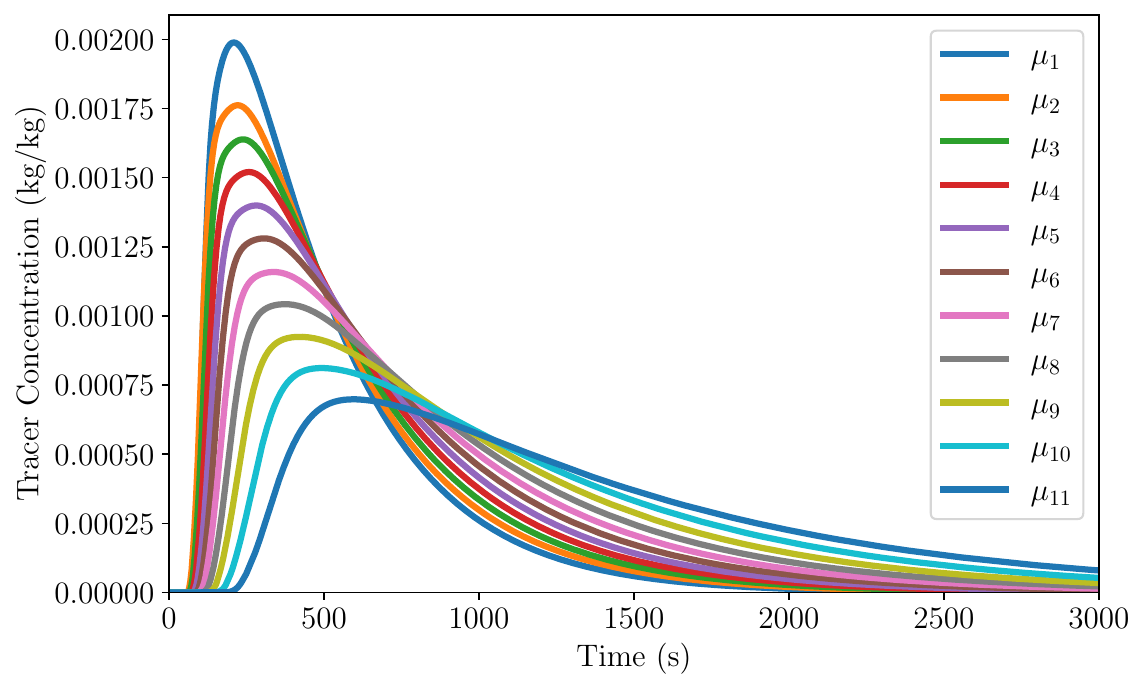}
    \caption{RTD curves obtained from full-order simulations, showing the temporal evolution of tracer concentration at the outlet of the tundish for each parameter listed in Table~\ref{Table:RTD_parameters}.}
    \label{fig:RTD_FOM_results}
\end{figure}

To construct the parametric data-driven ROM for RTD analysis (see Section~\ref{sec:ROM}), we consider parameter values corresponding to distinct simulation cases with varying inlet velocity conditions. The dataset consists of 11 parameter instances, of which 8 are selected randomly (from Table~\ref{Table:RTD_parameters}) for training, indexed as \( \boldsymbol{\mu}_i \) with \( i \in \{ 1, 2, 3, 4, 7, 8, 9, 10 \} \), while the remaining 3, \( i \in \{ 5, 6, 11 \} \), are reserved for testing. Full-order simulation snapshots corresponding to the training set are used to construct the reduced basis subspace. Further, nested POD is applied, and the first 21 modes capturing 99.99\% of the cumulative energy are retained. To approximate the mapping from the parameter-time space to the reduced coefficients, any regression technique discussed in Section~\ref{subsec:solmanifold} could be employed. In this work, radial basis function (RBF) interpolation is used. The test parameters are not used during training and serve to evaluate the predictive capability and generalization of the ROM, particularly in interpolation and mild extrapolation regimes.} \\

Figure~\ref{fig:EXP_FOM_ROM_prediction} compares the RTD curves for parameter $\mu_1$ obtained from the experiment (for details about the experiment, readers are referred to the previous work on intrusive projection-based ROM for RTD analysis \cite{Harshith2025PMOR}), the FOM CFD simulation, and the data-driven POD-RBF-ROM. Both the FOM and ROM predictions exhibit good agreement with the experimental data, with only minor discrepancies observed.  \\

\begin{figure}[ht!]
    \centering
    \includegraphics[width=0.72\linewidth]{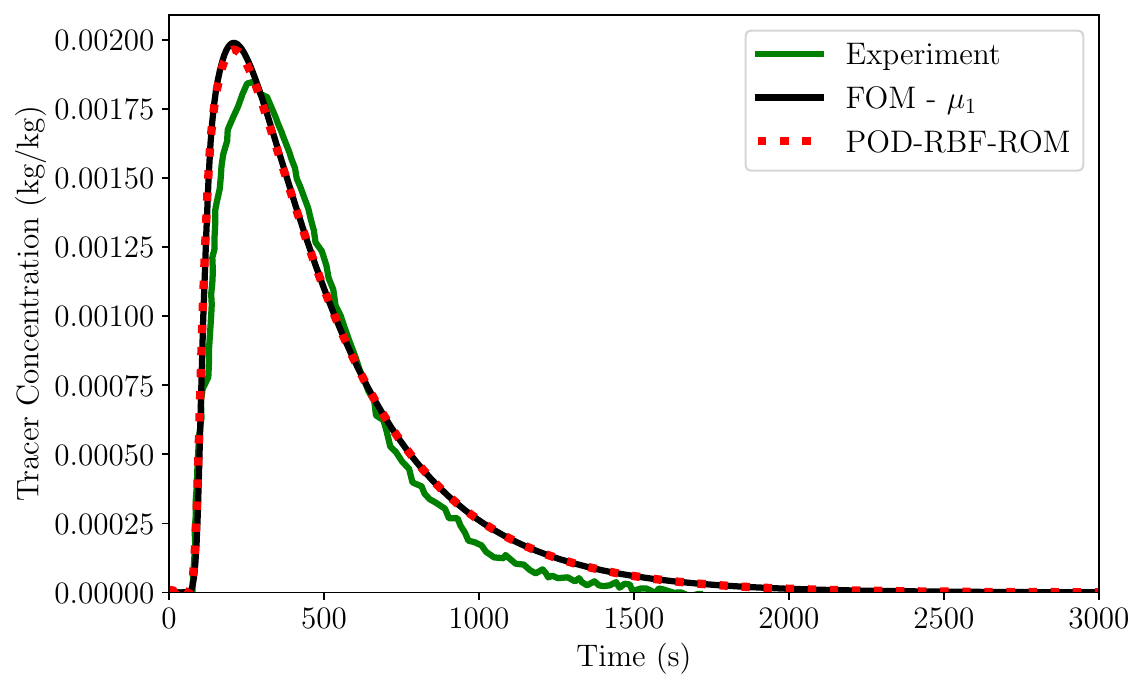}
    \caption{Comparison of RTD curves from the experiment \cite{Harshith2025PMOR}, FOM CFD simulation, and data-driven POD-RBF-ROM, illustrating the good agreement between the simulations and experimental data.}
    \label{fig:EXP_FOM_ROM_prediction}
\end{figure}

In the interpolatory regime, \(\boldsymbol{\mu}_i, \text{where}, i= \{5,6\}\), the data-driven POD-RBF-ROM shows excellent agreement with the full-order simulation, accurately predicting the RTD curves for test parameters within the range used for training. The model captures the dynamics of the system effectively, providing reliable results for these parameters. \\

\begin{figure}[ht!]
    \centering
    \includegraphics[width=0.7\linewidth]{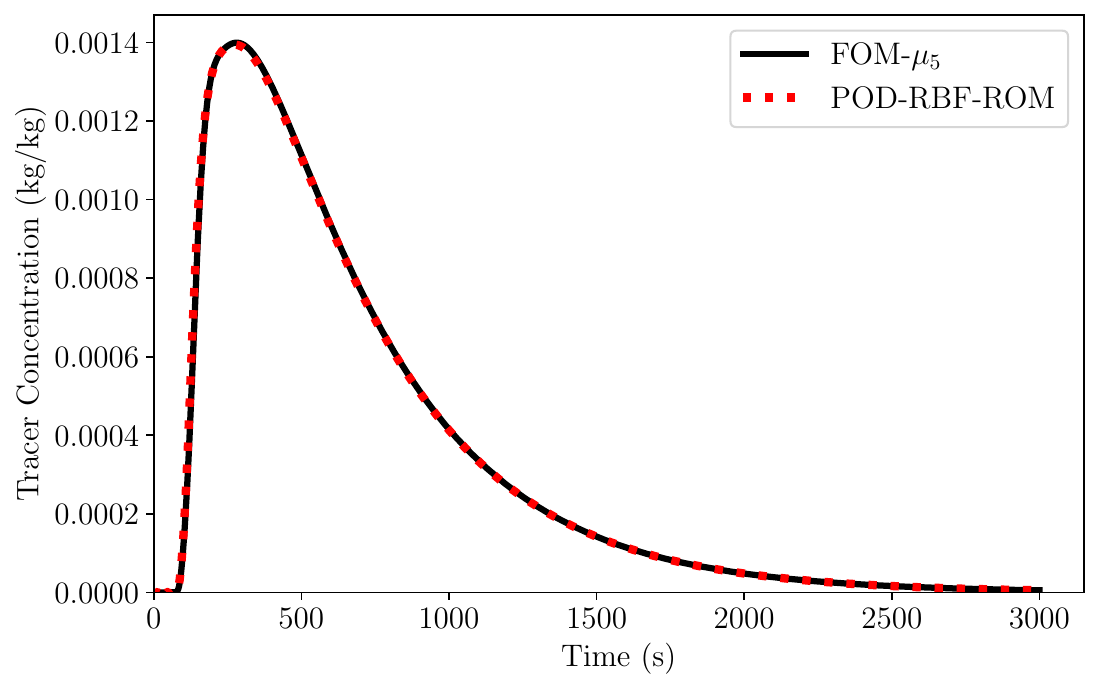}
    \caption{RTD curves for test parameter $\boldsymbol{\mu}_5$: comparison between the data-driven POD-RBF-ROM and full-order simulation, showing excellent agreement within the interpolatory regime}
    \label{fig:PMOR_prediction_1}
\end{figure}

\begin{figure}[ht!]
    \centering
    \includegraphics[width=0.7\linewidth]{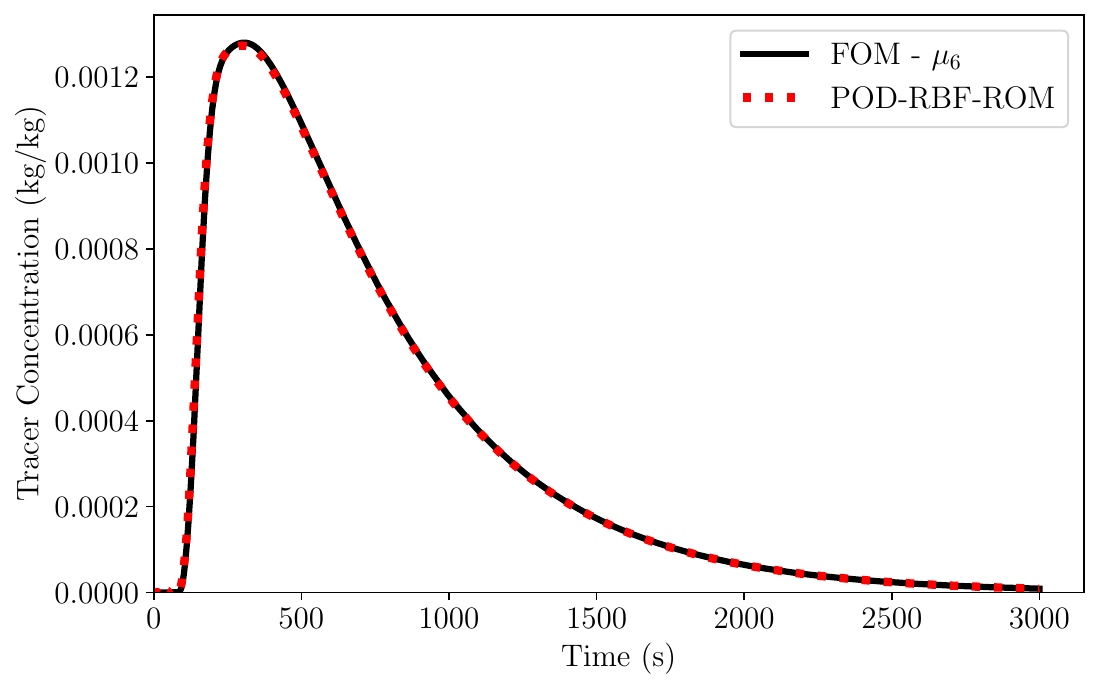}
    \caption{RTD curves for test parameter $\boldsymbol{\mu}_6$: comparison between the data-driven POD-RBF-ROM and full-order simulation, showing excellent agreement within the interpolatory regime}
    \label{fig:PMOR_prediction_2}
\end{figure}

In the extrapolatory regime, where the test parameter $\boldsymbol{\mu}_{11}$ is outside the training range, the ROM still captures the overall trend of the RTD curve but shows fluctuations in the initial phase and a slight discrepancy in concentration prediction. This demonstrates the ROM's ability to extrapolate beyond the trained parameters, though with some limitations in precision. \\

{ In Figure \ref{fig:relativeError}, we address the accuracy of ROM predictions for each parameter, the relative $L^2$ errors of the ROM predictions are low within the interpolation region, with training errors ranging from 1.0\% to 2.2\% (mean 1.7\%) and test errors between 1.7\% and 1.9\%, demonstrating accurate capture of system dynamics. In the extrapolation region, the test error rises to 8.9\%, indicating reduced generalization. These results highlight that while the ROM performs reliably within the training range, additional training data or more sophisticated regression strategy will improve extrapolation performance.} \\

\begin{figure}[ht!]
    \centering
    \includegraphics[width=0.7\linewidth]{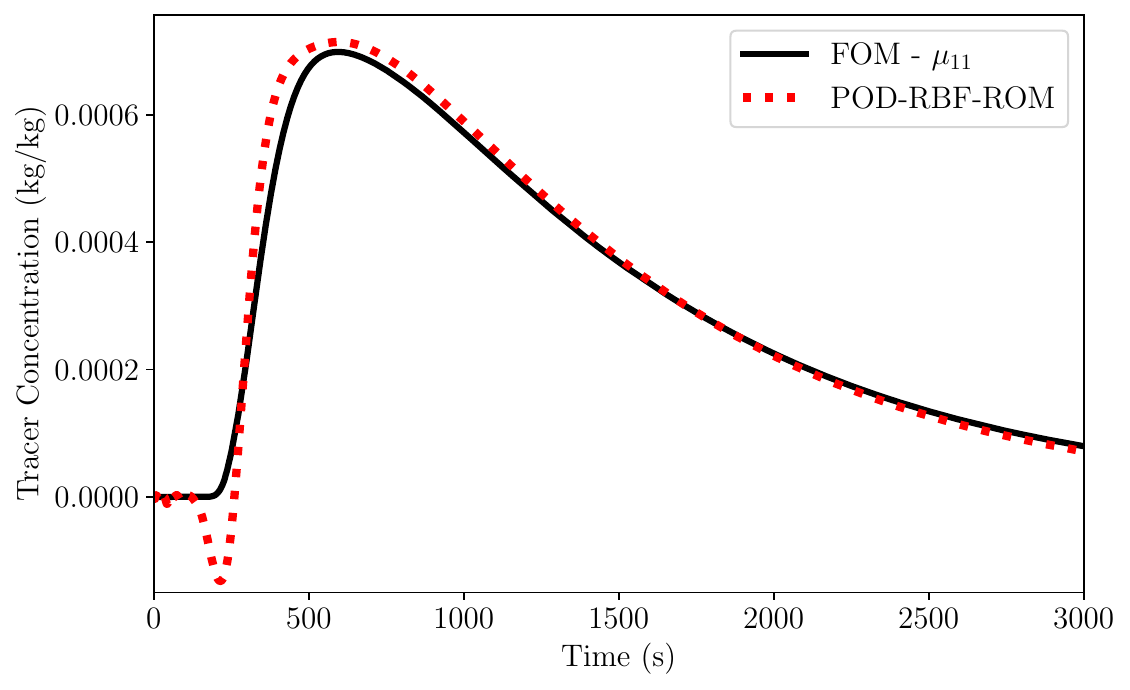}
    \caption{RTD curves for test parameter $\boldsymbol{\mu}_{11}$ in the extrapolatory regime: comparison between the data-driven POD-RBF-ROM and full-order simulation, ROM predictions showing fluctuations in the initial phase but accurately capturing the overall trend of the RTD curve, with a slight discrepancy in the concentration prediction.}
    \label{fig:PMOR_prediction_3}
\end{figure}

\begin{figure}
    \centering
    \includegraphics[width=0.75\linewidth]{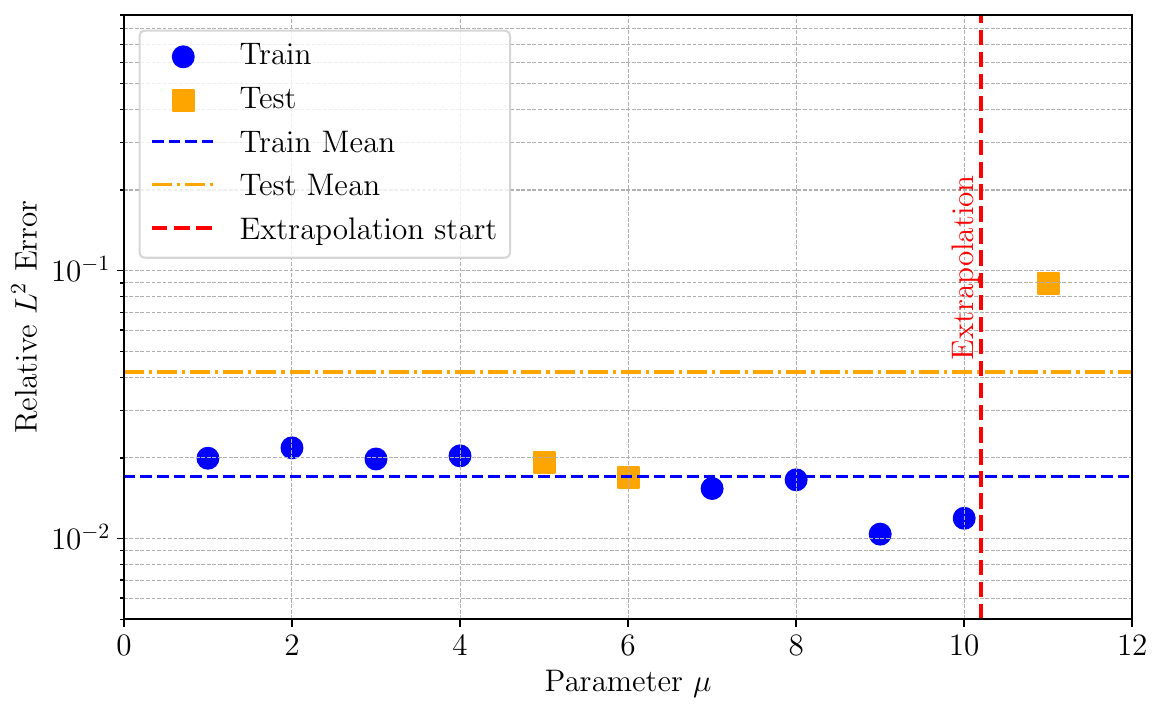}
    \caption{Relative $L_2$ errors of POD-RBF-ROM predictions. Left of the red line: interpolation region with low errors for both training (1–2\%) and test sets (1.7–1.9\%). Right of the red line: extrapolation region, where test error rises to 8.9\%.}
    \label{fig:relativeError}
\end{figure}

The computational resources and wall-clock time required for the FOM simulations are summarised in Table~\ref{tab:steady-state-FOMcomp-resources}. The computational efficiency achieved using the data-driven ROM for evaluating the QoIs is presented in Table~\ref{tab:steady-state-DD-ROMcomp-resources}. While individual FOM simulations range approximately from 8.19 and 23.17 hours of wall-clock time on 64 to 40 cores, the ROM’s online evaluation completes in only 0.33 seconds on a single core, resulting in a speed-up of approximately $1 \times {10 ^ 6}$ faster in wall-clock time compared to the average FOM simulation. Although the ROM offline construction demands around 40 hours on 2 cores, comparable to a single FOM simulation, this cost is a one-time investment, enabling rapid repeated evaluations. This substantial acceleration highlights the ROM’s suitability for real-time prediction and optimization tasks.

\begin{table}[ht!]
    \centering
    \renewcommand{\arraystretch}{1.2}
    \begin{tabular}{|>{\centering\arraybackslash}p{1.7cm}|>{\centering\arraybackslash}p{4.cm}|>{\centering\arraybackslash}p{4cm}|>{\centering\arraybackslash}p{2.8cm}|}        
    \hline
    \textbf{Expt No.} & \textbf{Wall-clock time} & \textbf{CPU time (core-hours)} & \textbf{Cores Allocated} \\     
    \hline
    1 &  2.81 + 5.38  hours &  177.05 + 360 hours    &   64 \\ 
    2 &  4.43 + 9 hours  &  177.2 + 360 hours   &   40 \\ 
    3 &  7.26 + 9 hours &  283.08 + 360 hours   &   40 \\ 
    4 &  6.67 + 9  hours &  260.92 + 360 hours   &   40 \\   
    5 &  5.16 + 9  hours &  201.55 + 360 hours   &   40 \\   
    6 &  4.24 + 9  hours &  165 + 360 hours     &   40 \\    
    7 &  14.17 + 9  hours &  555.24 + 360 hours   &   40 \\   
    8 &  3.75 + 9  hours &  145.53 + 360 hours   &   40 \\  
    9 &  6.10 + 9  hours &  237.70 + 360 hours   &   40 \\    
    10 &  6.54 + 9  hours &  256.27 + 360 hours   &   40 \\   
    11 &  10.48 + 9  hours &  410.90 + 360 hours   &   40 \\
    \hline
    \end{tabular}
    \caption{Summary of computational resources used for the FOM simulations to obtain the tundish steady-state operation QoIs on the Linux cluster. The wall-clock time refers to the elapsed real time, while the CPU time denotes the cumulative usage across all allocated cores (core-hours). These results correspond to the steady Reynolds averaged Navier-Stokes + transient tracer simulations discussed in Table \ref{Table:RTD_parameters}.}
    \label{tab:steady-state-FOMcomp-resources}
\end{table}

\begin{table}[ht!]
    \centering
    \renewcommand{\arraystretch}{1.2} 
    \begin{tabular}{|>{\centering\arraybackslash}p{3cm}|
                    >{\centering\arraybackslash}p{2.2cm}|
                    >{\centering\arraybackslash}p{2.5cm}|
                    >{\centering\arraybackslash}p{2.8cm}|
                    >{\centering\arraybackslash}p{2.cm}|}        
    \hline
    \textbf{Data-driven ROM} & \textbf{Offline Wall-clock Time} & \textbf{Offline CPU Time (core-hours)} & \textbf{Online Time (Wall-clock)} & \textbf{Cores Used (Offline / Online)} \\
    \hline
    POD-RBF-ROM &  40 hours &  80 core-hours &  0.33 seconds & 2 / 1 \\
    \hline
    \end{tabular}
    \caption{Computational resources used for the parametric time-dependent data-driven POD-RBF-ROM construction and prediction to obtain the tundish steady-state operation QoIs. The offline phase includes computation of POD modes (whole domain is considered), evaluation of modal coefficients, and RBF regression map construction. The online phase corresponds to the prediction of quantities of interest, the RTD curve at the tundish outlet. The online evaluation is performed only at the outlet probe location, where the QoI is recorded. Evaluating over the entire domain takes approximately 90 seconds, primarily due to the cost of the dot product operation.}
    \label{tab:steady-state-DD-ROMcomp-resources}
\end{table}

%% file: sections/conclusion.tex
\newpage

\section{Conclusions and perspectives}\label{sec:conclusion}

Over the last twenty years, mathematical modelling has become a critical approach for analysing flow behaviour in steelmaking tundishes. Given the difficulties associated with conducting experiments in high-temperature environments, physical studies have remained relatively constrained. However, advancements in computational methods have significantly enhanced our ability to study and optimise tundish performance through numerical simulations. \\

In this work, full-order simulations were conducted under both isothermal and non-isothermal conditions. The findings show that buoyancy effects associated with thermal variations have a negligible impact on the flow field, as the velocity profiles remain nearly identical between the two scenarios. These velocity fields, in conjunction with tracer transport simulations, serve as the foundation for evaluating residence time distribution and volume partitioning. Subsequently, a sensitivity analysis of the RTD curves was conducted by varying the inlet velocities. These results provide a thorough analysis of flow characteristics and mixing efficiency within the tundish. \\

A comparison of RTD curves obtained from experiments, full-order simulations, and the data-driven ROM indicates close agreement between the model predictions and experimental observations, with only minor discrepancies. This agreement validates the accuracy of the ROM and demonstrates its suitability as a computationally efficient surrogate for full-scale simulations in the analysis of tundish flow behaviour. In the interpolatory regime, the ROM exhibits excellent predictive capability, closely matching the FOM results. In the extrapolatory regime, while the ROM captures the overall shape of the RTD curve, some deviations are observed during the initial transient phase. \\

{ The proposed ROM achieves substantial computational acceleration, with online evaluations completing in approximately 0.33 seconds, corresponding to a speed-up of $10^6$ compared to full-order simulations. This demonstrates the strong potential of the ROM for real-time prediction of tundish RTD curves within the studied parameter range. While the current study is limited to a single-parameter scenario, these results provide a promising foundation for future extensions to higher-dimensional parametric spaces and broader digital twin applications.} \\

Future work will focus on extending the framework to more complex industrial scenarios, including multi-outlet tundish configurations. Further refinement is required in the extrapolatory regime, potentially by increasing the training data or using more sophisticated interpolation strategies.  \\

Overall, the integration of ROM techniques significantly reduces computational costs while retaining essential flow characteristics, demonstrating their potential for real-time analysis, design optimization, and digital twin application in metallurgical processes such as continuous casting.

%% file: sections/ackno.tex
\section*{\large CRediT authorship contribution statement}
\textbf{Harshith Gowrachari}: Writing - original draft, Conceptualization, Data curation, Formal Analysis, Investigation, Methodology, Software, Visualization. \textbf{Mattia Giuseppe Barra}: Writing – review $\&$ editing, Resources,  Investigation. \textbf{Giovanni Stabile}: Writing – review $\&$ editing, Supervision. \textbf{Gianluca Bazzaro}: Project administration, Supervision.   
\textbf{Gianluigi Rozza}:  Writing – review $\&$ editing, Funding acquisition, Project administration, Supervision.  

\section*{\large Declaration of Generative AI and AI-assisted technologies in the writing process}
These technologies were used to improve readability and correct spelling during the preparation of this manuscript. After using them, the authors reviewed and edited the content as needed and take full responsibility for the content of the publication. 

\section*{\large Declaration of competing interest}
The authors declare that they have no known competing financial interests or personal relationships that could have appeared to influence the work reported in this paper

\section*{\large Data statement}
Access to the data will be unavailable, as the research data includes sensitive and confidential information. 

\section*{Acknowledgements}
We acknowledge the PhD grant supported by industrial partner Danieli \& C. S.p.A. and Programma Operativo Nazionale Ricerca e Innovazione 2014-2020, P.I. Gianluigi Rozza. HG gratefully acknowledges Mattia Giuseppe Barra, Gianluca Bazzaro and Gabriele Guastaferro for their valuable discussions and for facilitating the hosting arrangements during multiple visits to Danieli Research Center (DRC).  GS acknowledges the financial support under the National Recovery and Resilience Plan (NRRP), Mission 4, Component 2, Investment 1.1, Call for tender No. 1409 published on 14.9.2022 by the Italian Ministry of University and Research (MUR), funded by the European Union – NextGenerationEU– Project Title ROMEU – CUP P2022FEZS3 - Grant Assignment Decree No. 1379 adopted on 01/09/2023 by the Italian Ministry of Ministry of University and Research (MUR) and acknowledges the financial support by the European Union (ERC, DANTE, GA-101115741). Views and opinions expressed are however those of the author(s) only and do not necessarily reflect those of the European Union or the European Research Council Executive Agency. Neither the European Union nor the granting authority can be held responsible for them. 

%% file: main_arxiv.bbl
\begin{thebibliography}{37}
\providecommand{\natexlab}[1]{#1}
\providecommand{\url}[1]{\texttt{#1}}
\expandafter\ifx\csname urlstyle\endcsname\relax
  \providecommand{\doi}[1]{doi: #1}\else
  \providecommand{\doi}{doi: \begingroup \urlstyle{rm}\Url}\fi

\bibitem[Anapagaddi et~al.(2013)Anapagaddi, Shukla, Mangal, and Singh]{anapagaddi2013exploring}
R.~Anapagaddi, R.~Shukla, S.~Mangal, and A.~K. Singh.
\newblock Exploring the design space of a continuous casting tundish for improved inclusion removal and reduced dead volume.
\newblock In \emph{Proceedings of the International Conference on Science and Technology of Ironmaking and Steelmaking, CSIR-National Metallurgical Laboratory, Jamshedpur}, 2013.

\bibitem[Anapagaddi et~al.(2014)Anapagaddi, Shukla, Goyal, Singh, Allen, Panchal, and Mistree]{Anapagaddi2014}
R.~Anapagaddi, R.~Shukla, S.~Goyal, A.~K. Singh, J.~K. Allen, J.~H. Panchal, and F.~Mistree.
\newblock Exploration of the design space in continuous casting tundish.
\newblock In \emph{Volume 2B: 40th Design Automation Conference}, IDETC-CIE2014. American Society of Mechanical Engineers, Aug. 2014.
\newblock \doi{10.1115/detc2014-34254}.
\newblock URL \url{http://dx.doi.org/10.1115/DETC2014-34254}.

\bibitem[Benner et~al.(2015)Benner, Gugercin, and Willcox]{Benner2015}
P.~Benner, S.~Gugercin, and K.~Willcox.
\newblock A survey of projection-based model reduction methods for parametric dynamical systems.
\newblock \emph{SIAM Review}, 57\penalty0 (4):\penalty0 483–531, Jan. 2015.
\newblock ISSN 1095-7200.
\newblock \doi{10.1137/130932715}.
\newblock URL \url{http://dx.doi.org/10.1137/130932715}.

\bibitem[Bensouici et~al.(2009)Bensouici, Bellaouar, and Talbi]{Bensouici2009}
M.~Bensouici, A.~Bellaouar, and K.~Talbi.
\newblock Numerical investigation of the fluid flow in continuous casting tundish using analysis of rtd curves.
\newblock \emph{Journal of Iron and Steel Research International}, 16\penalty0 (2):\penalty0 22–29, Feb. 2009.
\newblock ISSN 2210-3988.
\newblock \doi{10.1016/s1006-706x(09)60022-4}.
\newblock URL \url{http://dx.doi.org/10.1016/S1006-706X(09)60022-4}.

\bibitem[Bérard et~al.(2020)Bérard, Blais, and Patience]{Brard2020}
A.~Bérard, B.~Blais, and G.~S. Patience.
\newblock Experimental methods in chemical engineering: Residence time distribution—rtd.
\newblock \emph{The Canadian Journal of Chemical Engineering}, 98\penalty0 (4):\penalty0 848–867, Mar. 2020.
\newblock ISSN 1939-019X.
\newblock \doi{10.1002/cjce.23711}.
\newblock URL \url{http://dx.doi.org/10.1002/cjce.23711}.

\bibitem[Caretto et~al.(2007)Caretto, Gosman, Patankar, and Spalding]{Caretto}
L.~S. Caretto, A.~D. Gosman, S.~V. Patankar, and D.~B. Spalding.
\newblock \emph{Two calculation procedures for steady, three-dimensional flows with recirculation}, page 60–68.
\newblock Springer Berlin Heidelberg, 2007.
\newblock ISBN 9783540061717.
\newblock \doi{10.1007/bfb0112677}.
\newblock URL \url{http://dx.doi.org/10.1007/BFb0112677}.

\bibitem[Chen et~al.(2012)Chen, Cheng, Sun, Hou, Wang, and Zhang]{Chen2012}
C.~Chen, G.~Cheng, H.~Sun, Z.~Hou, X.~Wang, and J.~Zhang.
\newblock Effects of salt tracer amount, concentration and kind on the fluid flow behavior in a hydrodynamic model of continuous casting tundish.
\newblock \emph{steel research international}, 83\penalty0 (12):\penalty0 1141–1151, July 2012.
\newblock ISSN 1869-344X.
\newblock \doi{10.1002/srin.201200086}.
\newblock URL \url{http://dx.doi.org/10.1002/srin.201200086}.

\bibitem[Damle and Sahai(1995)]{Damle1995}
C.~Damle and Y.~Sahai.
\newblock The effect of tracer density on melt flow characterization in continuous casting tundishes. a modeling study.
\newblock \emph{ISIJ International}, 35\penalty0 (2):\penalty0 163–169, 1995.
\newblock ISSN 0915-1559.
\newblock \doi{10.2355/isijinternational.35.163}.
\newblock URL \url{http://dx.doi.org/10.2355/isijinternational.35.163}.

\bibitem[Dinda et~al.(2024)Dinda, Li, Guerra, Cathcart, and Barati]{Dinda2024}
S.~K. Dinda, D.~Li, F.~Guerra, C.~Cathcart, and M.~Barati.
\newblock Continuous casting tundish dead volume study by physical modeling and computational investigation.
\newblock \emph{steel research international}, 95\penalty0 (11), Sept. 2024.
\newblock ISSN 1869-344X.
\newblock \doi{10.1002/srin.202400125}.
\newblock URL \url{http://dx.doi.org/10.1002/srin.202400125}.

\bibitem[Gowrachari et~al.(2025)Gowrachari, Barra, Khamlich, Stabile, Bazzaro, and Rozza]{Harshith2025PMOR}
H.~Gowrachari, M.~G. Barra, M.~Khamlich, G.~Stabile, G.~Bazzaro, and G.~Rozza.
\newblock Projection-based model order reduction for residence time distribution analysis of an industrial-scale continuous casting tundish, 2025.
\newblock URL \url{https://arxiv.org/abs/2509.20366}.

\bibitem[Hesthaven and Ubbiali(2018)]{hesthaven2018non}
J.~Hesthaven and S.~Ubbiali.
\newblock Non-intrusive reduced order modeling of nonlinear problems using neural networks.
\newblock \emph{Journal of Computational Physics}, 363:\penalty0 55–78, June 2018.
\newblock ISSN 0021-9991.
\newblock \doi{10.1016/j.jcp.2018.02.037}.
\newblock URL \url{http://dx.doi.org/10.1016/j.jcp.2018.02.037}.

\bibitem[Hesthaven et~al.(2016)Hesthaven, Rozza, and Stamm]{hesthaven2016certified}
J.~S. Hesthaven, G.~Rozza, and B.~Stamm.
\newblock \emph{{Certified Reduced Basis Methods for Parametrized Partial Differential Equations}}.
\newblock Springer International Publishing, 2016.
\newblock ISBN 9783319224701.
\newblock \doi{10.1007/978-3-319-22470-1}.
\newblock URL \url{http://dx.doi.org/10.1007/978-3-319-22470-1}.

\bibitem[Launder and Spalding(1972)]{Launder1972}
B.~E. Launder and D.~B. Spalding.
\newblock \emph{Lectures in Mathematical Models of Turbulence}.
\newblock Academic Press, London and New York, 1972.
\newblock ISBN 0124380506.

\bibitem[Lee and Carlberg(2020)]{lee2020model}
K.~Lee and K.~T. Carlberg.
\newblock {Model reduction of dynamical systems on nonlinear manifolds using deep convolutional autoencoders}.
\newblock \emph{Journal of Computational Physics}, 404:\penalty0 108973, Mar. 2020.
\newblock ISSN 0021-9991.
\newblock \doi{10.1016/j.jcp.2019.108973}.
\newblock URL \url{http://dx.doi.org/10.1016/j.jcp.2019.108973}.

\bibitem[Liu et~al.(2022)Liu, Zhou, Zuo, Wu, and Wu]{Liu2022}
J.~Liu, P.~Zhou, X.~Zuo, D.~Wu, and D.~Wu.
\newblock Optimization of the liquid steel flow behavior in the tundish through water model experiment, numerical simulation and industrial trial.
\newblock \emph{Metals}, 12\penalty0 (9):\penalty0 1480, Sept. 2022.
\newblock ISSN 2075-4701.
\newblock \doi{10.3390/met12091480}.
\newblock URL \url{http://dx.doi.org/10.3390/met12091480}.

\bibitem[Quarteroni et~al.(2016)Quarteroni, Manzoni, and Negri]{quarteroni2015reduced}
A.~Quarteroni, A.~Manzoni, and F.~Negri.
\newblock \emph{Reduced Basis Methods for Partial Differential Equations}.
\newblock Springer International Publishing, 2016.
\newblock ISBN 9783319154312.
\newblock \doi{10.1007/978-3-319-15431-2}.
\newblock URL \url{http://dx.doi.org/10.1007/978-3-319-15431-2}.

\bibitem[Ren et~al.(2022)Ren, Yang, and Zhang]{Ren2022}
Y.~Ren, W.~Yang, and L.~Zhang.
\newblock Deformation of non-metallic inclusions in steel during rolling process: A review.
\newblock \emph{ISIJ International}, 62\penalty0 (11):\penalty0 2159–2171, Nov. 2022.
\newblock ISSN 1347-5460.
\newblock \doi{10.2355/isijinternational.isijint-2022-235}.
\newblock URL \url{http://dx.doi.org/10.2355/isijinternational.ISIJINT-2022-235}.

\bibitem[Romor et~al.(2023)Romor, Stabile, and Rozza]{romor2023non}
F.~Romor, G.~Stabile, and G.~Rozza.
\newblock {Non-linear Manifold Reduced-Order Models with Convolutional Autoencoders and Reduced Over-Collocation Method}.
\newblock \emph{Journal of Scientific Computing}, 94\penalty0 (3), Feb. 2023.
\newblock ISSN 1573-7691.
\newblock \doi{10.1007/s10915-023-02128-2}.
\newblock URL \url{http://dx.doi.org/10.1007/s10915-023-02128-2}.

\bibitem[Rozza et~al.(2022)Rozza, Stabile, and Ballarin]{rozza2022advanced}
G.~Rozza, G.~Stabile, and F.~Ballarin.
\newblock \emph{Advanced Reduced Order Methods and Applications in Computational Fluid Dynamics}.
\newblock Society for Industrial and Applied Mathematics, Jan 2022.
\newblock ISBN 9781611977257.
\newblock \doi{10.1137/1.9781611977257}.
\newblock URL \url{http://dx.doi.org/10.1137/1.9781611977257}.

\bibitem[Sahai and Ahuja(1986)]{Sahai1986}
Y.~Sahai and R.~Ahuja.
\newblock Steel flow and mixing of melt in steelmaking tundishes.
\newblock \emph{Ironmaking \& Steelmaking}, 13\penalty0 (5):\penalty0 241--247, 1986.

\bibitem[Sahai and Emi(1996)]{Sahai1996}
Y.~Sahai and T.~Emi.
\newblock Melt flow characterization in continuous casting tundishes.
\newblock \emph{ISIJ International}, 36\penalty0 (6):\penalty0 667–672, 1996.
\newblock ISSN 0915-1559.
\newblock \doi{10.2355/isijinternational.36.667}.
\newblock URL \url{http://dx.doi.org/10.2355/isijinternational.36.667}.

\bibitem[Sahai and Emi(2007)]{Sahai2007}
Y.~Sahai and T.~Emi.
\newblock \emph{Tundish Technology for Clean Steel Production}.
\newblock WORLD SCIENTIFIC, Dec. 2007.
\newblock ISBN 9789812790767.
\newblock \doi{10.1142/6426}.
\newblock URL \url{http://dx.doi.org/10.1142/6426}.

\bibitem[Sirovich(1987)]{Sirovich1987}
L.~Sirovich.
\newblock Turbulence and the dynamics of coherent structures. ii. symmetries and transformations.
\newblock \emph{Quarterly of Applied Mathematics}, 45\penalty0 (3):\penalty0 573–582, Oct. 1987.
\newblock ISSN 1552-4485.
\newblock \doi{10.1090/qam/910463}.
\newblock URL \url{http://dx.doi.org/10.1090/qam/910463}.

\bibitem[Stabile and Rozza(2018)]{stabile2018finite}
G.~Stabile and G.~Rozza.
\newblock Finite volume pod-galerkin stabilised reduced order methods for the parametrised incompressible navier–stokes equations.
\newblock \emph{Computers \& Fluids}, 173:\penalty0 273–284, Sept. 2018.
\newblock ISSN 0045-7930.
\newblock \doi{10.1016/j.compfluid.2018.01.035}.
\newblock URL \url{http://dx.doi.org/10.1016/j.compfluid.2018.01.035}.

\bibitem[Stabile et~al.(2019)Stabile, Ballarin, Zuccarino, and Rozza]{stabile2019reduced}
G.~Stabile, F.~Ballarin, G.~Zuccarino, and G.~Rozza.
\newblock A reduced order variational multiscale approach for turbulent flows.
\newblock \emph{Advances in Computational Mathematics}, 45\penalty0 (5–6):\penalty0 2349–2368, June 2019.
\newblock ISSN 1572-9044.
\newblock \doi{10.1007/s10444-019-09712-x}.
\newblock URL \url{http://dx.doi.org/10.1007/s10444-019-09712-x}.

\bibitem[Stabile et~al.(2025)]{ITHACA-FV}
G.~Stabile et~al.
\newblock {{ITHACA-FV: In real Time Highly Advanced Computational Applications for Finite Volumes - ROMs for OpenFOAM}}.
\newblock \url{https://github.com/ithaca-fv/ITHACA-FV}, 2025.

\bibitem[Taddei(2020)]{taddei2020registration}
T.~Taddei.
\newblock {A Registration Method for Model Order Reduction: Data Compression and Geometry Reduction}.
\newblock \emph{SIAM Journal on Scientific Computing}, 42\penalty0 (2):\penalty0 A997–A1027, Jan. 2020.
\newblock ISSN 1095-7197.
\newblock \doi{10.1137/19m1271270}.
\newblock URL \url{http://dx.doi.org/10.1137/19M1271270}.

\bibitem[Tezzele et~al.(2022{\natexlab{a}})Tezzele, Demo, Stabile, and Rozza]{Tezzele2022}
M.~Tezzele, N.~Demo, G.~Stabile, and G.~Rozza.
\newblock \emph{Chapter 9: Nonintrusive Data-Driven Reduced Order Models in Computational Fluid Dynamics}, page 203–222.
\newblock Society for Industrial and Applied Mathematics, Jan. 2022{\natexlab{a}}.
\newblock ISBN 9781611977257.
\newblock \doi{10.1137/1.9781611977257.ch9}.
\newblock URL \url{http://dx.doi.org/10.1137/1.9781611977257.ch9}.

\bibitem[Tezzele et~al.(2022{\natexlab{b}})Tezzele, Fabris, Sidari, Sicchiero, and Rozza]{Tezzele2022GPR}
M.~Tezzele, L.~Fabris, M.~Sidari, M.~Sicchiero, and G.~Rozza.
\newblock A multifidelity approach coupling parameter space reduction and nonintrusive pod with application to structural optimization of passenger ship hulls.
\newblock \emph{International Journal for Numerical Methods in Engineering}, 124\penalty0 (5):\penalty0 1193–1210, Nov. 2022{\natexlab{b}}.
\newblock ISSN 1097-0207.
\newblock \doi{10.1002/nme.7159}.
\newblock URL \url{http://dx.doi.org/10.1002/nme.7159}.

\bibitem[Tkadlečková et~al.(2020)Tkadlečková, Walek, Michalek, and Huczala]{Tkadlekov2020}
M.~Tkadlečková, J.~Walek, K.~Michalek, and T.~Huczala.
\newblock Numerical analysis of rtd curves and inclusions removal in a multi-strand asymmetric tundish with different configuration of impact pad.
\newblock \emph{Metals}, 10\penalty0 (7):\penalty0 849, June 2020.
\newblock ISSN 2075-4701.
\newblock \doi{10.3390/met10070849}.
\newblock URL \url{http://dx.doi.org/10.3390/met10070849}.

\bibitem[Wang et~al.(2021)Wang, Tan, Huang, Yan, Gu, He, and Li]{Wang2021}
Q.~Wang, C.~Tan, A.~Huang, W.~Yan, H.~Gu, Z.~He, and G.~Li.
\newblock Numerical simulation on refractory wear and inclusion formation in continuous casting tundish.
\newblock \emph{Metallurgical and Materials Transactions B}, 52\penalty0 (3):\penalty0 1344–1356, 2021.
\newblock ISSN 1543-1916.
\newblock \doi{10.1007/s11663-021-02097-7}.
\newblock URL \url{http://dx.doi.org/10.1007/s11663-021-02097-7}.

\bibitem[Wang et~al.(2022)Wang, Yang, Wang, Yue, Xia, and Xiao]{Wang2022}
Z.~Wang, Z.~Yang, X.~Wang, Q.~Yue, Z.~Xia, and H.~Xiao.
\newblock Residence time distribution (rtd) applications in continuous casting tundish: A review and new perspectives.
\newblock \emph{Metals}, 12\penalty0 (8):\penalty0 1366, Aug. 2022.
\newblock ISSN 2075-4701.
\newblock \doi{10.3390/met12081366}.
\newblock URL \url{http://dx.doi.org/10.3390/met12081366}.

\bibitem[Weller et~al.(1998)Weller, Tabor, Jasak, and Fureby]{Weller1998}
H.~G. Weller, G.~Tabor, H.~Jasak, and C.~Fureby.
\newblock A tensorial approach to computational continuum mechanics using object-oriented techniques.
\newblock \emph{Computers in Physics}, 12\penalty0 (6):\penalty0 620–631, Nov. 1998.
\newblock ISSN 0894-1866.
\newblock \doi{10.1063/1.168744}.
\newblock URL \url{http://dx.doi.org/10.1063/1.168744}.

\bibitem[{World Steel Association}(2024)]{worldsteel2024}
{World Steel Association}.
\newblock World steel in figures 2024, 2024.
\newblock URL \url{https://worldsteel.org/wp-content/uploads/World-Steel-in-Figures-2024.pdf}.

\bibitem[Xiao et~al.(2015)Xiao, Fang, Pain, and Hu]{Xiao2015}
D.~Xiao, F.~Fang, C.~Pain, and G.~Hu.
\newblock Non‐intrusive reduced‐order modelling of the navier–stokes equations based on rbf interpolation.
\newblock \emph{International Journal for Numerical Methods in Fluids}, 79\penalty0 (11):\penalty0 580–595, July 2015.
\newblock ISSN 1097-0363.
\newblock \doi{10.1002/fld.4066}.
\newblock URL \url{http://dx.doi.org/10.1002/fld.4066}.

\bibitem[Zhang and Thomas(2003)]{Zhang2003}
L.~Zhang and B.~G. Thomas.
\newblock Inclusions in continuous casting of steel.
\newblock In \emph{Proceedings of the XXIV National Steelmaking Symposium}, pages 184--198, Morelia, Michoacán, Mexico, November 26--28 2003.
\newblock URL \url{https://ccc.illinois.edu/PDF%20Files/Presentations/03_Inclusions_in_Continuous_Casting_of_Steel.pdf}.

\bibitem[Zhang et~al.(2020)Zhang, Ren, Duan, Ren, Chen, Cheng, Yang, and Sridhar]{Zhang2020}
L.~Zhang, Q.~Ren, H.~Duan, Y.~Ren, W.~Chen, G.~Cheng, W.~Yang, and S.~Sridhar.
\newblock Modelling of non-metallic inclusions in steel.
\newblock \emph{Mineral Processing and Extractive Metallurgy}, 129\penalty0 (2):\penalty0 184–206, Mar. 2020.
\newblock ISSN 2572-665X.
\newblock \doi{10.1080/25726641.2020.1738087}.
\newblock URL \url{http://dx.doi.org/10.1080/25726641.2020.1738087}.

\end{thebibliography}
